\documentclass[aps,rmp,twocolumn,groupaddress]{revtex4-1}

\usepackage{amsmath}
\usepackage{amssymb}
\usepackage{bm}
\usepackage{url}
\usepackage[dvipdfmx]{graphicx}
\usepackage{color}

\makeatletter
\def\l@subsubsection#1#2{}
\makeatother

\begin{document}

\title{Topological Insulators}

\author{Yoichi Ando}
\email{ando@ph2.uni-koeln.de}
\affiliation{Physics Institute II, University of Cologne, Z\"ulpicher Str. 77, 50937 Cologne, Germany}

\begin{abstract}

Topological insulators are characterized by insulating bulk and conducting surface, the latter is a necessity consequence of the nontrivial topology of the wavefunctions forming the valence band. This chapter gives a historical overview of the discovery of topological insulators and a concise description of the ${\bm Z}_2$ topology which defines them. The concept of topological insulators have been extended to various other topologies, giving rise to the recognition of further topological states of matter such as topological crystalline insulators and higher-order topological insulators.
Representative materials of topological insulators, their synthesis techniques, and the ways for the experimental confirmation of the topological nature are introduced. Among the interesting phenomena derived from topological insulators, topological superconductivity, Majorana zero modes, and quantum anomalous Hall effect are briefly discussed.

\end{abstract}

\maketitle

%\tableofcontents

\begin{flushleft}
\end{flushleft}

\begin{flushleft}
{\bf Key Points}\\
$\bullet$ Bulk-boundary correspondence guarantees the appearance of gapless boundary states in a topological material \\
$\bullet$ Topological insulators are defined by the ${\bm Z}_2$ topology that was theoretically discovered in 2005 \\
$\bullet$ Spin-momentum locking is a key characteristic of the boundary states of topological insulators, allowing for topological superconductivity to show up in proximity to a conventional superconductor \\
$\bullet$ Most topological-insulator materials are not really insulating in the bulk, but bulk-insulating samples can be obtained for Bi$_2$Te$_2$Se,  Bi$_{2-x}$Sb$_x$Te$_{3}$, Bi$_{2-x}$Sb$_x$Te$_{3-y}$Se$_y$, HgTe, and WeTe$_2$.\\
$\bullet$ Quantum anomalous Hall effect can be realized in ferromagnetic topological insulators
\end{flushleft}

%%%% INTRODUCTION
\section{Introduction}

Topological insulators (TIs) are a class of insulators whose occupied states are characterized by a nontrivial topological invariant. In the widest sense of the term, a two-dimensional (2D) electron gas in high magnetic fields showing the quantum Hall effect can be considered a TI, because such a system is characterized by a non-zero Chern number, which is a prototypical topological invariant for a time-reversal-breaking system. However, the term TI is most commonly used for time-reversal-invariant insulators that are characterized by a non-zero ${\bm Z}_2$ index. Both 2D and three-dimensional (3D) systems can be TIs, and a 2D TI may also be called a quantum spin Hall insulator for historical reasons. While a 2D TI is characterized by a single ${\bm Z}_2$ index, a combination of four ${\bm Z}_2$ indices is necessary to fully characterize 3D TIs, which can be subdivided into strong TIs and weak TIs based on the ${\bm Z}_2$ indices. New types of TIs characterized by different topological invariants, such as topological crystalline insulators or higher-order TIs, have been discovered with concrete materials realizations.

An important consequence of the nontrivial topology of TIs is that a gapless boundary state necessarily shows up when the material is physically terminated. This is because the nontrivial topology is a discrete characteristic of the gapped energy states, which cannot change as long as the energy gap remains open. Since the vacuum (i.e. outside of the TI) is also gapped and is topologically trivial, there is a collision of different topologies at the boundary. In order for the topology to change across the boundary, the gap must close there. Therefore, 3D TIs are always associated with gapless surface states, and so are two-dimensional 2D TIs with gapless edge states. Furthermore, the resulting gapless states always reflect the topological character of the bulk states, which is called bulk-boundary correspondence. 

In 2D and 3D TIs, their nontrivial topology is protected by time-reversal symmetry (TRS). There are special points in the Brillouin zone called time-reversal-invariant momenta, at which the momentum $\mathbf{k}$ is invariant upon time reversal. The Kramers theorem dictates that the gapless boundary states of a TI must be degenerate at a time-reversal-invariant momentum; as a result, the energy dispersion of the boundary states of a TI takes the form of a massless Dirac cone located at one of the time-reversal-invariant momenta. This massless Dirac cone is spin-non-degenerate, and the requirement of TRS means that the direction of the spin must be opposite between the states with $\mathbf{k}$ and $\mathbf{-k}$. This relationship is called spin-momentum locking (see Fig. 1), which is one of the most important characteristics of TIs. In a 2D TI, the spin-momentum locking in the 1D edge state leads to the quantum spin Hall effect. 
When the TRS is broken, the gapless (or massless) nature of the Dirac cone is no longer protected by the topology and a gap may open at the Dirac point, but it does not mean that the Dirac cone disappears altogether upon TRS breaking.

\begin{figure}
\begin{center}
\includegraphics[width=3cm]{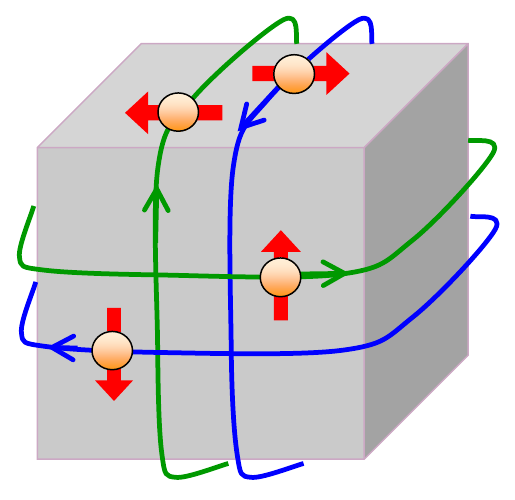}
\caption{Schematic picture of the spin-momentum locking in the topological surface states of a 3D TI. Electrons moving in opposite directions (depicted by green or blue lines with arrows) have opposite spin directions (red arrows); note that spins are perpendicular to the momentum.}
\label{fig:HelicalSS}
\end{center}
\end{figure}

\section{Historical Perspective}

The quantum Hall effect (QHE) was discovered in 1980 by von Klitzing {\it et al.} in a high-mobility 2D semiconductor under high magnetic fields \cite{Klitzing1980}. In 1982, Thouless, Kohmoto, Nightingale, and den Nijs (TKNN) \cite{Thouless1982} calculated the Hall conductance $\sigma_{xy}$ of a Landau-quantized 2D system and showed that, when the chemical potential lies in a gap between Landau levels, $\sigma_{xy} = \nu(e^2/h)$ with an integer $\nu$. This integer $\nu$, often called TKNN invariant, was later shown to be nothing but the first Chern number of a 2D TRS-breaking topological system \cite{Kohmoto1985}. Therefore, the importance of topology in condensed matter was already recognized in the QHE system in 1980s.

Another important event preceding TIs was the discovery in 2004 of the spin Hall effect \cite{Kato2004}, the appearance of transverse spin current in response to longitudinal electric field. The spin Hall effect in nonmagnetic systems is fundamentally related to the anomalous Hall effect in ferromagnets \cite{Nagaosa2010} and it can have both intrinsic and extrinsic origins. The intrinsic mechanism stems from the Berry curvature of the valence-band Bloch wavefunctions integrated over the Brillouin zone. Since such an integral can be non-zero even in an insulator, Murakami, Nagaosa, and Zhang proposed the idea of spin Hall insulator \cite{Murakami2004}, which is a gapped insulator with a finite spin Hall conductivity.
Although this idea was later shown to be unrealistic as a bulk property \cite{Onoda2005}, this triggered the subsequent proposal of the quantum spin Hall insulator in 2005 by Kane and Mele \cite{Kane2005a, Kane2005b}, who considered the edge states associated with an insulating 2D bulk. The ingenious proposal by Kane and Mele provided a concrete model (graphene with spin-orbit coupling) to realize the bulk band gap and gapless edge states in zero magnetic field \cite{Kane2005a}.
More importantly, Kane and Mele recognized that the electronic states of their quantum spin Hall insulator is characterized by a novel topology specified by a ${\bm Z}_2$ index \cite{Kane2005b}. This theoretical discovery of the ${\bm Z}_2$ topology in insulators was an important step in our understanding of topological phases of matter. Experimentally, since the spin-orbit coupling in graphene is very weak, it is unfortunately difficult to observe the quantum spin Hall effect predicted in the Kane-Mele model. However, in 2006 Bernevig, Hughes, and Zhang proposed another 2D model to produce a ${\bm Z}_2$ topological phase based on the band structure of HgTe, and they predicted that a CdTe/HgTe/CdTe quantum well should be a quantum spin Hall insulator \cite{Bernevig2006}. This prediction was verified in 2007 by K\"onig {\it et al.} \cite{Konig2007}, who observed the quantization of the longitudinal conductance to $2e^2/h$ in zero magnetic field, giving evidence for the gapless edge states in the insulting regime. 

In parallel with the experimental verification of the ${\bm Z}_2$ topology in 2D, the topological classification of time-reversal-invariant insulators was extended to 3D systems. It was shown in 2006-2007 that four ${\bm Z}_2$ indices are needed to fully characterize the topology for 3D \cite{Moore2007, Fu2007, Roy2009}. The term ``topological insulator" was coined by Moore and Balents when they proposed the existence of topological systems in 3D \cite{Moore2007}. Fu and Kane made a concrete prediction that the Bi$_{1-x}$Sb$_x$ alloy in the insulating composition should be a 3D TI, and they further proposed that the nontrivial topology can be verified with the angle-resolved photoemission spectroscopy (ARPES) by counting the number of times the surface states cross the Fermi energy between two time-reversal-invariant momenta \cite{Fu2007b}. The proposed experiment was conducted by Hsieh {\it et al.} who reported in 2008 that Bi$_{1-x}$Sb$_x$ is indeed a 3D TI \cite{Hsieh2008}. 

Before the ${\bm Z}_2$ topology was theoretically discovered by Kane and Mele, a time-reversal-invariant topological state was conceived for four-dimensional (4D) systems, for which the topological invariant is an integer \cite{Zhang2001}. The effective field theory constructed for such 4D topological systems \cite{Bernevig2002} was shown to be useful for 3D and 2D TIs \cite{Qi2008}, leading to interesting predictions regarding topological magnetoelectric effects \cite{Qi2008}.

\section{Theory of topological insulators}

The Berry connection of the Bloch wavefunctions in the momentum space plays the key role in the theory of topological insulators. 
Let us consider a band insulator described by a Bloch Hamiltonian $H({\bm k})$ with the crystal momentum ${\bm k}$. The eigenstates are given by the solutions of the Bloch equation,
\begin{eqnarray}
H({\bm k})|u_n({\bm k})\rangle=E_n({\bm k})|u_n({\bm k})\rangle.
\label{eq2:eigen}
\end{eqnarray}
The Berry connection $\bm{\mathcal A}^{(n)}({\bm k})$ is defined by
\begin{eqnarray}
\bm{\mathcal A}^{(n)}({\bm k})=i\langle u_n({\bm k})|{\bm \partial}_{{\bm
k}}u_n({\bm k})\rangle,
\label{eq2:Berryconnection}
\end{eqnarray}
which measures the rate of change in the wavefunction $|u_n({\bm k})\rangle$ in the
momentum space. 
%The rotation of the Berry connection is the Berry curvature $\bm{\mathcal B}^{(n)}({\bm k})$:
%\begin{eqnarray}
%\bm{\mathcal B}^{(n)}({\bm k}) = \nabla_{\bm k} \times \bm{\mathcal A}^{(n)}({\bm k}).
%\label{eq2:Berrycurvature}
%\end{eqnarray}

A gauge-invariant quantity constructed from $\bm{\mathcal A}^{(n)}({\bm k})$ is the Berry phase, which is defined by
\begin{eqnarray}
\gamma_n[C] \equiv \oint_C d{\bm k}\cdot\bm{\mathcal A}^{(n)}({\bm k}),
\label{eq2:integral}
\end{eqnarray}
where the line integral is performed along a closed path $C$ in the momentum space and the gauge is chosen so that $\bm{\mathcal A}^{(n)}({\bm k})$ is non-singular on $C$.
Note that $\gamma_n[C]$ has a freedom of $2\pi N$ with integer $N$ and is not strictly gauge-invariant, but $e^{i\gamma_{n}[C]}$ is.

Another gauge-invariant quantity that can be constructed from $\bm{\mathcal A}^{(n)}({\bm k})$ is the field strength defined by
\begin{eqnarray}
{\cal F}^{(n)}_{ij}({\bm k}) \equiv \partial_{k_i}{\cal A}^{(n)}_{k_j}({\bm k})
-\partial_{k_j}{\cal A}^{(n)}_{k_i}({\bm k}). 
\end{eqnarray}
For a 2D system, the Chern number of the $n$-th band is defined as the integral of the field strength ${\cal F}^{(n)}_{ij}({\bm k})$ over the 2D Brillouin zone:
\begin{eqnarray}
Ch^{(n)} =\frac{1}{2\pi}\int_{\rm 2D\,BZ} dk_x dk_y {\cal F}^{(n)}_{xy}({\bm k}).
\label{eq2:Ch1}
\end{eqnarray}
This integral vanishes if $\bm{\mathcal A}^{(n)}({\bm k})$ has no singularity over the Brillouin zone.
But if $\bm{\mathcal A}^{(n)}({\bm k})$ has a singularity at ${\bm k}_0$, the gauge transformation
\begin{eqnarray}
\bm{\mathcal A}^{(n)'}({\bm k}) 
=\bm{\mathcal A}^{(n)}({\bm k})-{\bm \partial}_{\bm k}\phi_n({\bm k})
\end{eqnarray}
in the region $R$ including ${\bm k}_0$ can make $\bm{\mathcal A}^{(n)'}({\bm k})$ to have no singularity in $R$. 
Using the Stokes' theorem, Eq. (\ref{eq2:Ch1}) becomes
\begin{eqnarray}
Ch^{(n)} = \frac{1}{2\pi}\int_{\partial R} d{\bm k}\cdot {\bm \partial}_{\bm
k}\phi_n({\bm k}),
\end{eqnarray}
where $\partial R$ is the boundary of $R$.
Because $e^{i\phi_n({\bm k})}$ is a unique function on $\partial R$, this $Ch^{(n)}$ must be an integer.
For a 2D insulator, the total Chern number $Ch$ of the occupied bands is given by
\begin{eqnarray}
Ch=\sum_{E_n<E_{\rm F}}Ch^{(n)}.
\label{eq3:totalCh}
\end{eqnarray} 
Note that $Ch$ can be finite only when the time-reversal symmetry is broken, because $Ch = -Ch$ under time-reversal symmetry.

% Z2 index

The key insight of Kane and Mele \cite{Kane2005b} was that the Kramers degeneracy makes it possible to define a new topological invariant (i.e. the ${\bm Z}_2$ index) for 2D time-reversal-invariant systems. The time-reversal operator ${\cal T}$ obeys ${\cal T}^2=-1$ for spin-$\frac{1}{2}$ electrons, and the anti-unitarity of ${\cal T}$ yields 
\begin{eqnarray}
\langle {\cal T} u | {\cal T} v\rangle=\langle v| u \rangle 
\end{eqnarray}
for any states $|u\rangle$ and $|v\rangle$.
This leads to the conclusion that a state $|u \rangle$ is orthogonal to its time-reversal partner ${\cal T}|u\rangle$, i.e. $\langle u|{\cal T}u\rangle=0$, because $\langle u|{\cal T}u\rangle=\langle {\cal T}^2 u|{\cal T}u\rangle=-\langle u|{\cal T} u\rangle$.
In a time-reversal-invariant system, $|u\rangle$ and ${\cal T} |u\rangle$ have the same energy, and hence their orthogonality means a guaranteed two-fold degeneracy, which is the essence of the Kramers degeneracy.

The time-reversal invariance makes the Bloch Hamiltonian $H({\bm k})$ in Eq. (\ref{eq2:eigen}) to have the following property:
\begin{eqnarray}
{\cal T} H({\bm k}){\cal T}^{-1}=H(-{\bm k}). \label{eq:Bloch-TRS}
\end{eqnarray} 
Hence, for an eigenstate $|u_n({\bm k})\rangle$ of $H({\bm k})$, its Kramers partner ${\cal T}|u_n({\bm k})\rangle$ is an eigenstate of $H(-{\bm k})$. The orthogonality between $|u_n({\bm k})\rangle$ and ${\cal T}|u_n({\bm k})\rangle$ means that the latter can also be written by using a different eigenstate $|u_{n'}({\bm k})\rangle$ of $H({\bm k})$  as $|u_{n'}(-{\bm k})\rangle$. Hereafter, we write such a Kramers pair $(|u_n({\bm k})\rangle, |u_{n'}({\bm k})\rangle)$ as $(|u_n^{\rm I}({\bm k})\rangle, |u^{\rm II}_{n}({\bm k})\rangle)$. Due to the phase ambiguity of the eigenstates, there is a general relation
\begin{eqnarray}
|u_n^{\rm II}({\bm k})\rangle=e^{i\varphi_n({\bm k})}{\cal T}|u^{\rm I}_n(-{\bm
k})\rangle, 
\end{eqnarray}
with $\varphi_n({\bm k})$ the gauge degree of freedom.

The ${\bm Z}_2$ index was originally provided in the form of a Pfaffian \cite{Kane2005b}, but here we derive it as a variant of the spin Chern number \cite{Sheng2006}, which is well-defined when a 2D system has a fixed spin orientation, say $S_z$. In the diagonal basis of $S_z$, the Bloch Hamiltonian is written as 
\begin{eqnarray}
H({\bm k})=\left(
\begin{array}{cc}
H_{\uparrow}({\bm k}) & 0\\
0 & H_{\downarrow}({\bm k})
\end{array}
\right) ,
\end{eqnarray}
where $H_{\uparrow}({\bm k})$ and $H_{\downarrow}({\bm k})$) are for the $S_z=1/2$ and $S_z=-1/2$ sector, respectively. With this separation of $H({\bm k})$, the spin Chern number $Ch_{\uparrow}$ and $Ch_{\downarrow}$ are defined as the Chern number of $H_{\uparrow}({\bm k})$ and $H_{\downarrow}({\bm k})$, respectively. Since each spin sector breaks time-reversal symmetry, the spin Chern number can be nonzero even for a time-reversal-invariant system, although the total Chern number must always vanish (i.e. $Ch_{\uparrow}+Ch_{\downarrow}=0$) due to time-reversal symmetry.

In general, spin-orbit coupling breaks spin conservation, and thus the spin Chern number is not well-defined in a spin-orbit coupled system. However, in the presence of time-reversal symmetry, one can derive an analogous topological number as follows: Instead of the spin eigensectors, we use Kramers pairs $(|u_n^{\rm I}({\bm k})\rangle, |u_n^{\rm II}({\bm k})\rangle)$ to divide the Hilbert space into two subspaces, and then introduce the Chern numbers in those subspaces,
\begin{eqnarray} 
&&Ch_{\rm I}=\frac{1}{2\pi}\int_{\rm 2D\,BZ} dk_xdk_y
{\cal F}^{{\rm I}(-)}_{xy}({\bm k}), 
\nonumber\\
&&Ch_{\rm II}=\frac{1}{2\pi}\int_{\rm 2D\,BZ} dk_xdk_y
{\cal F}^{{\rm II}(-)}_{xy}({\bm k}), 
\end{eqnarray}
where ${\cal F}^{{\rm I}(-)}_{xy}({\bm k})$ and ${\cal F}^{{\rm II}(-)}_{xy}({\bm k})$ are the field strengths of the Berry curvatures $\bm{\mathcal A}^{{{\rm I}(-)}}({\bm k})$ and $\bm{\mathcal A}^{{{\rm II}(-)}}({\bm k})$, which are defined for the occupied states as 
\begin{eqnarray}
&&\bm{\mathcal A}^{{{\rm I}(-)}}({\bm k})=i\sum_{E_n<E_{\rm F}}
\langle u_n^{\rm I}({\bm k})|\partial_{\bm k} u_n^{\rm I}({\bm k})\rangle, 
\nonumber\\
&&\bm{\mathcal A}^{{{\rm II}(-)}}({\bm k})=i\sum_{E_n<E_{\rm F}}
\langle u_n^{\rm II}({\bm k})|\partial_{\bm k} u_n^{\rm II}({\bm k})
\rangle. 
\end{eqnarray}
These Chern numbers share the same property as the spin Chern numbers: they take integer numbers, and $Ch_{\rm I}+Ch_{\rm II}=0$ due to time-reversal symmetry. However, $Ch_{\rm I}$ and $Ch_{\rm II}$ themselves are not really well-defined, because the superscripts ${\rm I}$ and ${\rm II}$ do not have any physical meaning, and hence they can be exchanged; such an exchange leads to $Ch_{\rm I}\rightarrow Ch_{\rm II}$ (=$-Ch_{\rm I}$) and $Ch_{\rm II}\rightarrow Ch_{\rm I}$ (=$-Ch_{\rm II}$), and thus the Chern numbers $Ch_{\rm I}$ and $Ch_{\rm II}$ cannot be unique. 

Nevertheless, by considering only the parity of these Chern numbers, i.e. $(-1)^{Ch_{\rm I}}$ and $(-1)^{Ch_{\rm II}}$, this ambiguity can be avoided; namely, because of the constraint $Ch_{\rm I}=-Ch_{\rm II}$, the parity satisfies $(-1)^{Ch_{\rm I}} = (-1)^{-Ch_{\rm I}} = (-1)^{Ch_{\rm II}}$, making them robust against the exchange of the superscripts. As a result, we have a unique ${\bm Z}_2$ index $(-1)^{\nu_{\rm 2D}}$ which is defined as
\begin{eqnarray}
(-1)^{\nu_{\rm 2D}}\equiv (-1)^{Ch_{\rm I}}=(-1)^{Ch_{\rm II}}. 
\label{eq2:2dZ2}
\end{eqnarray}
An insulator with $(-1)^{\nu_{\rm 2D}}=-1$ is topologically distinct from an ordinary insulator and such a system is called a 2D TI or a quantum spin Hall insulator. $\nu_{\rm 2D}$ is called ${\bm Z}_2$ topological invariant and takes the value 0 or 1.

\begin{figure}[tb]
\centering
\includegraphics[width=0.5\columnwidth]{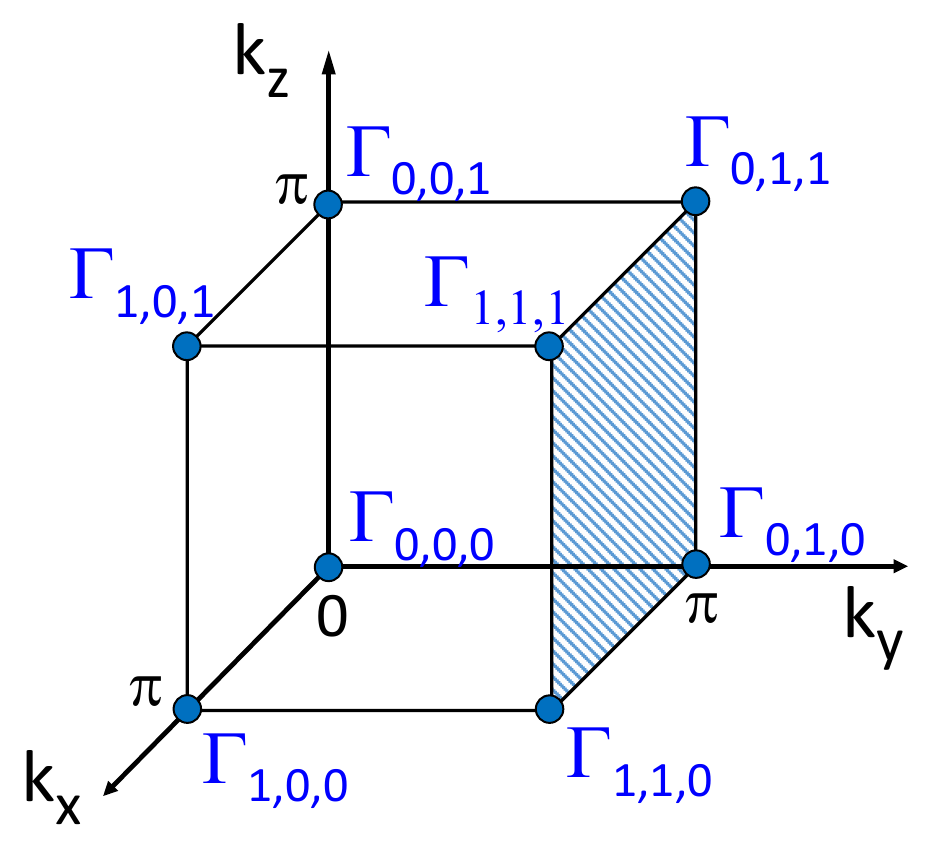}
\caption
{The eight time-reversal-invariant momenta $\Gamma_i$ $[i = (n_1, n_2, n_3)]$ in the 3D Brillouin zone having tetragonal symmetry. The time-reversal-invariant plane $\Sigma^a_{n_a}$ for $a$ = 2 and $n_a$ = 1 is hatched as an example.
} 
\label{fig:bz}
\end{figure}

The time-reversal invariance allows us to introduce ${\bm Z}_2$ indices in three dimensions \cite{Moore2007}. A time-reversal-invariant momentum $\Gamma_i$ is defined as a momentum-space point satisfying $-\Gamma_i=\Gamma_i+{\bm G}$ for a reciprocal lattice vector ${\bm G}$.  In a 3D Brillouin zone, there are a total of eight such $\Gamma_i$'s and they can be indexed by using three integers $n_a$ ($a$ =1, 2, 3) taking the value of 0 or 1 as 
\begin{eqnarray}
\Gamma_{i=(n_1,n_2,n_3)}=\frac{1}{2}
\left(n_1 {\bm b}_1+n_2{\bm b}_2+n_3{\bm b}_3\right) 
\end{eqnarray}
with the primitive reciprocal lattice vectors ${\bm b}_i$ (see Fig. \ref{fig:bz}). Fixing one of the three $n_a$'s gives a group of four time-reversal-invariant momenta. For example, by fixing $n_2$ to be 1, one obtains four time-reversal-invariant momenta $\Gamma_{(n_1,1,n_3)}$ with $n_1=0,1$ and $n_3=0,1$. These four time-reversal-invariant momenta specifies a distinct time-reversal-invariant plane $\Sigma^a_{n_a}$ in the 3D Brillouin zone (see Fig. \ref{fig:bz} for the example of $\Sigma^a_{n_a}$ for $a$ = 2 and $n_a$ = 1). One can easily see that there are six distinct time-reversal-invariant planes in the 3D Brillouin zone.

It is easy to see that each time-reversal-invariant plane $\Sigma^a_{n_a}$ has its own ${\bm Z}_2$ index $(-1)^{Ch_{\rm I}(\Sigma_{n_a}^a)}$. Hence, there are a total of six ${\bm Z}_2$ indices that are defined in this manner, but they are not completely independent and the following constraints exist \cite{Moore2007}: 
\begin{eqnarray}
(-1)^{Ch_{\rm I}(\Sigma_{1}^1)-Ch_{\rm I}(\Sigma_{0}^1)} 
&=&(-1)^{Ch_{\rm I}(\Sigma_{1}^2)-Ch_{\rm I}(\Sigma_{0}^2)} 
\nonumber\\
&=&(-1)^{Ch_{\rm I}(\Sigma_{1}^3)-Ch_{\rm I}(\Sigma_{0}^3)}.
\end{eqnarray}
As a result, only the following four ${\bm Z}_2$ indices are independent \cite{Moore2007}: 
\begin{eqnarray}
&&(-1)^{\nu_{{\rm 2D},a}}\equiv
(-1)^{Ch_{\rm I}(\Sigma_1^a)}, \quad (a=1,2,3)
\nonumber\\
&&(-1)^{\nu_{\rm 3D}}\equiv
(-1)^{Ch_{\rm I}(\Sigma_{1}^1)-Ch_{\rm I}(\Sigma_{0}^1)}
= (-1)^{Ch_{\rm I}(\Sigma_{1}^1)}(-1)^{Ch_{\rm I}(\Sigma_{0}^1)}. \nonumber\\
\label{eq2:3dZ2}
\end{eqnarray}
The first three ${\bm Z}_2$ indices becomes nontrivial (i.e. $-1$) if the system is just a stack of 2D TIs, and these three are called weak indices. On the other hand, the last ${\bm Z}_2$ index $(-1)^{\nu_{\rm 3D}}$ is intrinsic to 3D and this is called strong ${\bm Z}_2$ index. An insulator with nontrivial strong ${\bm Z}_2$ index is called a strong 3D TI, while those having only nontrivial weak indices are called weak 3D TIs. The set of four parameters $(\nu_{\rm 3D}; \nu_{{\rm 2D},1}, \nu_{{\rm 2D},2}, \nu_{{\rm 2D},3})$ constitutes the 3D ${\bm Z}_2$ topological invariants to fully characterize the ${\bm Z}_2$ topology of a 3D insulator.

When an insulator has inversion symmetry, the evaluation of the ${\bm Z}_2$ indices becomes simple \cite{Fu2007b}. The inversion operator $P$ has the property $P^2=1$ and the Bloch Hamiltonian $H({\bm k})$ of a centrosymmetric system satisfies 
\begin{eqnarray}
PH({\bm k})P^{-1}=H(-{\bm k}).
\end{eqnarray}
From this relation, one can conclude that an eigenstate of $H({\bm k})$ at ${\bm k}=\Gamma_i$ is simultaneously an eigenstate of $P$. Furthermore, because of the relation $[{\cal T}, P]=0$, a Kramers pair have a common eigenvalue $\xi_i=\pm 1$ of $P$:
\begin{eqnarray}
&&P |u_n^{\rm I}(\Gamma_i)\rangle=\xi_i |u_n^{\rm I}(\Gamma_i)\rangle, 
\nonumber\\
&&P |u_n^{\rm II}(\Gamma_i)\rangle=\xi_i |u_n^{\rm II}(\Gamma_i)\rangle.
\end{eqnarray}
The ${\bm Z}_2$ index in 2D is evaluated by using these $\xi_i$ as 
\begin{eqnarray}
(-1)^{\nu_{\rm 2D}}=\prod_{i}\xi_i, 
\end{eqnarray} 
where the product is taken for all inequivalent $\Gamma_i$'s in 2D (there are 4 of them). 
The ${\bm Z}_2$ indices for 3D systems are given by
\begin{eqnarray}
(-1)^{\nu_{{\rm 2D},a}}=\prod_{n_a=1;n_{b\neq a}}\xi_{i=(n_1n_2n_3)} 
\end{eqnarray} 
and 
\begin{eqnarray}
(-1)^{\nu_{\rm 3D}}=\prod_i\xi_i \,, 
\label{eq2:3dparity_formula}
\end{eqnarray}
where the product for the strong index is taken for all eight $\Gamma_i$'s in the 3D Brillouin zone.

\section{Model Hamiltonian of a 3D topological insulator}
\label{sec:TI}

The nontrivial electronic structure of TIs stems from strong spin-orbit coupling. As an example, here we consider a prototypical topological insulator, Bi$_2$Se$_3$. Its bulk bands near the Fermi energy are mainly derived from the $p_z$ orbitals of two inequivalent Se atoms in the unit cell, which makes the effective Hamiltonian to have a $4 \times 4$ matrix form \cite{Zhang2009}
\begin{eqnarray}
H_{\rm TI}({\bm k})=(m_0-m_1{\bm k}^2)\sigma_x+v_zk_z\sigma_y +v\sigma_z(k_x s_y-k_y s_x),
\nonumber\\ 
\label{eq2:HamiltonianTI}
\end{eqnarray} 
with $\sigma_i$ denoting the Pauli matrix in the orbital space and $s_i$ the Pauli matrix in the spin space. Note that the eigenstates of $\sigma_z$ (i.e. $\sigma_z=\pm 1$) correspond to the two $p_z$ orbitals. The third term is a result of strong spin-orbit coupling. Since the inversion operation exchanges the $p_z$ orbitals, one can see that the inversion operator $P$ is given by $\sigma_x$. The $H_{\rm TI}({\bm k})$ is invariant under the inversion $P$, i.e.
\begin{eqnarray}
PH_{\rm TI}(-{\bm k})P^{-1}=H_{\rm TI}({\bm k}). 
\label{eq2:inversion_BiSe}
\end{eqnarray} 
The model Hamiltonian Eq. (\ref{eq2:HamiltonianTI}) focuses on the physics near the $\Gamma$ point (i.e. ${\bm k}$ = 0), and the strong ${\bm Z}_2$ index given by Eq. (\ref{eq2:3dparity_formula}) can be evaluated by taking the multiple of the parity of the occupied states at ${\bm k}$ = 0 and at the Brillouin zone boundary ($k \rightarrow \infty$ in this model).

Let us now consider the case $m_0 > 0$ and $m_1 > 0$ and examine if the system is topological. At ${\bm k}$ = 0 we have $H_{\rm TI}({\bm k} =0) = m_0 \sigma_x$, and hence the eigenstate with odd parity ($\sigma_x = -1$) becomes the occupied state when $m_0 > 0$. On the other hand, since we have $H_{\rm TI}({\bm k} \rightarrow \infty) \sim -m_1 k^2 \sigma_x$ at large $k$, the even-parity eigenstate becomes occupied when $m_1 > 0$. Thus, in this case, the strong ${\bm Z}_2$ index (i.e. the multiple of the parity of the occupied states at ${\bm k}$ = 0 and at $k \rightarrow \infty$) is $-1$, meaning that the system is topological. More generally, one can infer that the system becomes topological when $m_0$ and $m_1$ have the same sign, which is expressed as
\begin{eqnarray}
(-1)^{\nu_{\rm 3D}}=-{\rm sgn}\left(m_0m_1\right).
\end{eqnarray}
Physically, the condition for the system to become topological is that the parity of the occupied state switches at somewhere in the Brillouin zone between ${\bm k}$ = 0 and $k \rightarrow \infty$; such a switching is called band inversion, and the above discussion shows that it is the band inversion which makes Bi$_2$Se$_3$ to be topological.

Let us now derive the topological surface states from $H_{\rm TI}({\bm k})$. We consider a surface at $z=0$ with the insulator occupying the positive $z$ region. The wavefunction at the surface has the form
\begin{eqnarray}
\psi_{k_x, k_y}(z)=\left(e^{-\kappa_-z}-e^{-\kappa_+z}\right)
\left(
\begin{array}{c}
0 \\
1
\end{array}
\right)_{\sigma}
\otimes u_s(k_x, k_y), 
\nonumber\\
\end{eqnarray} 
with 
\begin{eqnarray}
\kappa_{\pm}=-\frac{v_z}{2m_1}\pm\sqrt{\left(\frac{v_z}{2m_1}\right)^2+k_x^2+k_y^2-\frac{m_0}{m_1}}. 
\end{eqnarray} 
Substituting $\psi_{k_x, k_y}(z)$ into the Schr\"{o}dinger equation 
\begin{eqnarray}
H_{\rm TI}(k_x, k_y, -i\partial_z)\psi_{k_x, k_y}(z)=E\,\psi_{k_x, k_y}(z) ,
\end{eqnarray}
one obtains a 2D Dirac equation 
\begin{eqnarray}
-v(k_x s_y-k_y s_x)u_s(k_x, k_y)=E\,u_s(k_x, k_y). 
\end{eqnarray}
The term $k_x s_y-k_y s_x$ dictates the spin-momentum locking of the 2D Dirac fermions; the spin is perpendicular to the surface momentum ${\bm k}_{\rm 2D}$ and lies within the surface plane.

\section{Representative topological insulator materials}

Usually, the TI materials become topological due to a band inversion, as one sees in the example  in the previous section.
The ultra-thin layer of HgTe is the most studied 2D TI. In the bulk HgTe, an inversion between $p$- and $s$-orbital bands gives the system a potential to be a TI, but these two bands are degenerate at the $\Gamma$ point. The quantum-confinement effect in a thin layer opens a gap and makes the system to be a 2D TI \cite{Bernevig2006}. Applying an epitaxial strain to a thick layer of HgTe can also open a gap and make the system to be a 3D TI \cite{Bruene2011}. Both TI phases have been experimentally confirmed. The highest mobility for the topological surface states of a 3D TI ($\sim 3 \times 10^4$ cm$^2$/Vs) has been obtained in strained HgTe, although the bulk band gap attainable in strained HgTe would be limited to $\sim$22 meV \cite{Bruene2011}. The HgTe system has not gained a wide popularity in the TI research, because the growth of HgTe films requires elaborate facilities.

The quantum well of AlSb/InAs/GaSb/AlSb realizing an inverted band alignment was theoretically predicted \cite{Liu2008} and experimentally confirmed \cite{Knez2011} to be a 2D TI. The valence-band top of GaSb lies above the conduction-band bottom of InAs, so when thin layers of InAs and GaSb are in direct contact and they are both quantum confined, the resulting hole subband in GaSb may lie above the electron subband in InAs, realizing the band inversion necessary for a TI. However, it has been reported that the 2D TI phase in this system is very fragile against disorder and it is difficult to utilize the edge transport in this system \cite{Shojaei2018}.

The monolayer of WTe$_2$ has recently emerged as a promising 2D TI \cite{Qian2014}. WTe$_2$ is a van-der-Waals material and a monolayer flake can be obtained by exfoliation. It was shown that the quantized conductance due to the 1D edge states persists up to $\sim$100 K in WTe$_2$ \cite{Wu2018}. 

The first material that was confirmed \cite{Hsieh2008} to be a 3D TI, Bi$_{1-x}$Sb$_x$, has a relatively high 2D carrier mobility of $\sim$10$^4$ cm$^2$/Vs \cite{Taskin2009}. Also, it is relatively easy for this system to reduce the bulk carrier density to the 10$^{16}$ cm$^{-3}$ level in high-quality single crystals, making it possible, for example, to perform Landau level spectroscopy of the surface states via magneto-optics \cite{Schafgans2012}. The bulk band gap is up to $\sim$30 meV, which is large enough to detect the 2D transport properties at 4 K. 

The most widely used 3D TIs for various studies of the topological phenomena are the family of materials having the tetradymite structure, Bi$_2$Se$_3$, Bi$_2$Te$_3$, Sb$_2$Te$_3$, and their relatives. Their 3D TI nature was predicted in 2009 \cite{Zhang2009}, followed by experimental confirmations of the existence of the topological surface states \cite{Xia2009, Chen2009}. Their crystal structure consists of covalently bonded quintuple layers (QLs), e.g. Se-Bi-Se-Bi-Se, that are stacked and weakly bonded by the van der Waals force, making it easily to cleave between QLs. High-quality single crystals and thin films of the binary compounds (Bi$_2$Se$_3$, Bi$_2$Te$_3$, Sb$_2$Te$_3$) can be relatively easily grown, which is one of the reasons for the popularity of these 3D TIs. It should be noted that those binary compounds are always conducting in the bulk and they are not really insulators, because bulk carrier doping due to naturally-occurring defects is unavoidable for entropy reasons. In 2010, the ternary compound Bi$_2$Te$_2$Se was discovered to be the first bulk-insulating 3D TI material \cite{Ren2010}, where the concealment of Se in the middle of the QL plays a key role in achieving bulk insulation. It is also possible to synthesize bulk-insulating samples by employing the concept of compensation, namely, balancing donors and acceptors so that they cancel each other. A series of compositions to realize bulk-insulation have been identified for Bi$_{2-x}$Sb$_x$Te$_{3-y}$Se$_y$ which can be grown as high-quality bulk single crystals \cite{Ren2011}; for example, Bi$_{1.5}$Sb$_{0.5}$Te$_{1.7}$Se$_{1.3}$ shows a particularly low bulk conductivity \cite{Taskin2011}. Also, it has been shown that bulk-insulating thin films can be obtained by optimizing the growth condition of Bi$_{2-x}$Sb$_x$Te$_{3}$ \cite{Zhang2011a}.

A less studied materials family of 3D TIs is TlBiSe$_2$ and its variants (TlBiTe$_2$, TlSbTe$_2$, TlBiSe$_{2-x}$S$_x$, etc.). In the solid solution TlBiSe$_{2-x}$S$_x$, one can access the topological phase transition between the TI and non-TI phases by tuning the $x$ value \cite{Sato2011, Xu2011}. Bulk-insulating single crystals can be grown by tuning the composition in TlBi$_x$Sb$_{1-x}$Te$_2$ \cite{Breunig2017}, in which the compensation similar to that in Bi$_{2-x}$Sb$_x$Te$_{3}$ can be achieved.

An interesting material in the context of 3D TIs is SmB$_6$, which may be a topological Kondo insulator in which the band gap originates from strong electron correlations \cite{Dzero2010}. There are strong evidence for surface-dominated transport at low temperature, but its topological nature remains controvertial. Note that the surface-dominated transport can be observed in trivial insulators due to the formation of an accumulation or inversion layer, as reported for pure Te \cite{Klitzing1971}.

%An interesting family of materials in the context of 3D TIs is (PbSe)$_5$(Bi$_2$Se$_3$)$_{3m}$ compound ($m$ = 1, 2, 3, 4) \cite{Nakayama,Segawa}. They form a natural multilayer heterostructure consisting of a TI (Bi$_2$Se$_3$) and an ordinary insulator (PbSe), in which the PbSe unit provides a barrier for the electronic states in the Bi$_2$Se$_3$ unit to be quantum confined, leading to hybridization of the topological interface states in each Bi$_2$Se$_3$ unit. This results in a gapped Dirac cone , resulting in a gapped Dirac-cone as a bulk state. 

\section{Topological crystalline insulators}

The ${\bm Z}_2$ topology protected by time-reversal symmetry is not the only possible topological classification of band insulators. Another topology protected by point-group symmetries of the crystal lattice can also classify insulators, and a material that is nontrivial with such a topology is called a topological crystalline insulator (TCI) \cite{Fu2011}. Concrete topological invariants have been constructed for systems possessing four-fold ($C_4$) or six-fold ($C_6$) rotation symmetry \cite{Fu2011} and also for systems having mirror symmetry \cite{Hsieh2012}. 

In this context, SnTe is a TCI protected by mirror symmetry, even though it is trivial in the ${\bm Z}_2$ topology. Here, the topology is specified by the topological invariant called mirror Chern number $n_{\mathcal{M}}$, which evaluates the Chern number in only one of the two Hilbert subspaces divided according to the mirror eigenvalues \cite{Teo2008}. SnTe is characterized by $n_{\mathcal{M}} = -2$, and its surface states on the \{001\} surface present a peculiar double-Dirac-cone structure near the $\bar{X}$ points, which was theoretically predicted \cite{Hsieh2012} and experimentally confirmed \cite{Tanaka2012}. Interestingly, the surface states on the \{111\} surface have a different structure consisting of two different Dirac cones at $\bar{\Gamma}$ and $\bar{M}$ points \cite{Tanaka2013a}. The TCI phase has also been reported for Pb$_{0.77}$Sn$_{0.23}$Se, in which a transition to a topologically-trivial phase was observed upon increasing temperature \cite{Dziawa2012}.  The topological phase transition also occurs as a function of the composition in Pb$_{1-x}$Sn$_x$Te, in which the TCI phase is realized for $x_c \gtrsim 0.25$ \cite{Xu2012, Tanaka2013b}. 

The mirror Chern number $n_{\mathcal{M}}$ can also be used for time-reversal-invariant 3D TIs to further classify them \cite{Teo2008}. For example, Bi$_{1-x}$Sb$_x$ is a TI with $Z_2$ invariant (1;111), and it can have $n_{\mathcal{M}} = \pm 1$. The sign of $n_{\mathcal{M}}$ is called mirror chirality, which is related to the sign of the $g$ factor. It was experimentally elucidated that Bi$_{1-x}$Sb$_x$ has the mirror chirality $-1$ \cite{Nishide2010}.

\section{Higher-order topological insulators}

It has been shown that 3D insulators can have a topology that leaves the surface to be gapped but some of their hinges or corners to accommodate symmetry-protected gapless states. The topological corner states may have fractional charge \cite{Benalcazar2017}. Such topological materials are called higher-order topological insulators --- a 2nd-order 3D TI has gapless hinge states, and a 3rd-order 3D TI or a 2nd-order 2D TI has gapless corner states \cite{Schindler2018NP, Schindler2018SA, Benalcazar2017}.
For example, in the presence of time-reversal and mirror symmetries, a 2nd-order 3D TI with helical hinge modes can be conceived, and SnTe has been proposed to be such a higher-order TI \cite{Schindler2018SA}. Recently, Bi was elucidated to be a 2nd-order 3D TI in which the topology is protected by three-fold rotation and inversion symmetries, with the experimental proof of the existence of helical hinge states on the (111) surface \cite{Schindler2018NP}.

\section{Experimental confirmation of TI\lowercase{s}}

In the case of 2D TIs, the existence of helical 1D edge states should be probed, which is possible only through quantum transport experiments using device structures. The existence of edge states can be confirmed through conductance quantization when the Fermi level is gate-tuned into the bulk band gap \cite{Konig2007}. Also, the helical spin polarization of the edge states may be detected by transport experiments using spin Hall effect \cite{Bruene2012}.

For 3D TIs, observation of the Dirac-cone surface states by ARPES experiments is the most straightforward. To firm up the identification of a TI, spin-resolved ARPES should be employed to detect the helical spin polarization of the Dirac cone \cite{Hsieh2009, Nishide2010}.
In transport experiments, it is necessary to prepare a sufficiently bulk-insulating sample to observe Shubnikov-de Haas (SdH) oscillation coming from high-mobility surface carriers, but if it is successful, the 2D Dirac nature of the topological surface states can be elucidated by the dependence of the SdH-oscillation frequency on the direction of the applied magnetic fields as well as by the $\pi$ Berry phase that can be identified via the Landau-level fan-diagram analysis \cite{Taskin2011b}.
The 2D Dirac nature may also be confirmed by scanning tunneling epectroscopy (STS) experiments in magnetic fields, through the observation of the peculiar Landau quantization in which the level spacing changes as $\sqrt{N}$ and the zero-energy Landau level is pinned to the Dirac point \cite{Chen2010}.

\section{Syntheses of TI Materials}

The synthesis of bulk single crystals of the chalcogenide 3D TI materials (Bi$_2$Se$_3$, Bi$_2$Te$_2$Se, SnTe, etc.) are done in sealed evacuated quartz-glass tubes, because chalcogen atoms are volatile. The necessity of containment
limits the range of applicable growth techniques, and the Bridgman method or vapor transport method are usually employed. In the Bridgman method, stoichiometric amounts of raw materials are first melted and then gradually cooled down in the presence of a temperature gradient, so that the crystallization starts at the cold end of the tube and the crystal grows as the solidification proceeds from this end. In the vapor transport technique, a chunk of polycrystalline material is put on the hot end of a sealed quartz-glass tube and the tube is kept for a long time in a furnace with a certain temperature gradient, during which the polycrystalline material sublimates and crystallizes at somewhere in the colder part. 

For the growth of high-quality epitaxial thin films of TI materials, molecular beam epitaxy (MBE) technique is usually employed, in which the constituent elements are co-evaporated with suitable flux ratios. Although the lattice matching between the substrate and the grown material is usually very important for an epitaxial growth, in the case of the TI materials having the tetradymite structure, the lattice matching with the substrate is not crucial and the epitaxial growth proceeds in the so-called van-der-Waals epitaxy mode. For example, epitaxial growths of Bi$_2$Se$_3$ is possible on various substrates including Si(111), graphene-terminated 6H-SiC(0001), SrTiO$_3$(111), GaAs(111), sapphire(0001), and InP(111).

Nanowires and nanoribbons of 3D TI materials are of interest because of the peculiar quantum-confined 1D states realized in those structures \cite{Breunig2021}. Nanowires of Bi$_2$Te$_3$ or similar compounds are usually synthesized by gold-catalyzed vapor liquid solid (VLS) technique \cite{Peng2010}. In this technique, heated Au nanoparticles (typically $\sim$20-nm diameter) are exposed to a flow of the vaporized material, which is absorbed at the top of the melted Au nanoparticles (liquid) and crystalizes at their bottom (solid), resulting in a forest of standing nanowires having a Au nanoparicle at their top.

\section{Interesting phenomena derived from TI\lowercase{s}}

\subsection{Topological superconductivity and Majorana zero modes}

The spin-momentum locking in the topological surface states of 3D TIs makes them a promising platform to realize topological superconductivity hosting Majorana zero modes (MZMs), which are emergent zero-energy states described by a real creation operator $\gamma$. 
When the creation operator $\gamma$ of a particle is real, it is self-conjugate, i.e. $\gamma = \gamma^{\dagger}$, and there is no distinction between creation and annihilation of the particle; in such a situation, one can say that the particle is its own antiparticle, which is the key characteristics of a Majorana particle and $\gamma$ is called Majorana operator. Importantly, a fermion creation operator $f^{\dagger}$ can be constructed by using two Majorana operators as $f^{\dagger} = (\gamma_1 + i \gamma_2)/2$. One can easily see that the number operator of this fermion is written as $f^{\dagger}f = (1- i\gamma_1 \gamma_2)/2$, which means that $-i\gamma_1 \gamma_2$ is 1 ($-1$) when the fermionic state is occupied (unoccupied). This operator $-i\gamma_1 \gamma_2$ is called parity operator, and it signifies the occupancy of the zero-energy fermionic state $f$ by an electron. This situation allows to encode quantum information in terms of the fermion parity by using two MZMs, which is the basis of topological quantum computation \cite{Nayak2008}. 

%When a particle is its own antiparticle, such a particle is said to have the Majorana character. if a particle is created by the operator $\gamma$, its antiparticle is created by $\gamma^{\dagger}$, and hence the Majorana character means $\gamma = \gamma^{\dagger}$, i.e. self-conjugate. This condition is naturally satisfied if $\gamma$ is a real operator. With such a $\gamma$, one can construct a fermion creation operator $f^{\dagger}$ by using two Majorana creation operators as  $f^{\dagger} = (\gamma_1 + i \gamma_2)/2$. One can easily see that the number operator of the fermion is written as $f^{\dagger}f = (1- i\gamma_1 \gamma_2)/2$, which means that $-i\gamma_1 \gamma_2$ is 1 ($-1$) when the fermionic state is occupied (unoccupied). This operator $-i\gamma_1 \gamma_2$ is called parity operator, and it signifies the occupancy of the zero-energy fermionic state $f$ by an electron. This allows to encode the quantum information in terms of parity by using two MZMs, which is the basis of topological quantum computation. 

Fu and Kane showed \cite{Fu2008} that when superconductivity is induced in the topological surface states by using the superconducting proximity effect from a conventional $s$-wave superconductor in contact with a 3D TI, the resulting 2D superconducting state is topological and a MZM will shop up in the vortex core. In usual $s$-wave superconductors, the spin-singlet Cooper pairs always leads to spin degeneracy in the in-gap states created in the superconducting condensate, and the spin degeneracy prohibits in-gap states to have the Majorana character. In the topological surface states, however, the spin-momentum locking removes this degeneracy, and further breaking of time-reversal symmetry in the vortex core leads to the MZM. Nanowires of 3D TIs are also an interesting platform to host MZMs. In this case, the quantum confinement of the topological surface states leads to spin-degenerate 1D subbands, but their degeneracy can be lifted by a parallel magnetic field to recover the spin-momentum locking. When superconductivity is induced in such subbands, 1D topological superconductivity is realized and MZMs will show up a the ends of the nanowire \cite{Cook2011}. Further breaking inversion symmetry by electrostatic gating is predicted to give rise to stable MZMs \cite{Legg2021}.

\subsection{Quantum anomalous Hall effect}

The quantum anomalous Hall effect (QAHE) occurs without Landau quantization and it signifies the Chern number of the occupied states of a 2D ferromagnetic insulator. The surface states of 3D TIs can be gapped when time-reversal symmetry is broken by a ferromagnetic order, which can be induced in 3D TIs by doping magnetic elements. When such a ferromagnetic 3D TI is made into an ultra-thin film (thickness $\lesssim$ 10 nm) and the Fermi level is tuned into the gap opened at the Dirac point, the QAHE can be realized. 
In the QAHE, the electrical current is carried by the spin-polarized 1D chiral edge state, which causes the Hall conductance $\sigma_{yx}$ to be quantized to $e^2/h$ and the longitudinal conductance $\sigma_{xx}$ to vanish. 
Experimentally, the QAHE has been observed in the bulk-insulating TI material (Bi$_{1-x}$Sb$_x$)$_2$Te$_3$ doped with Cr or V \cite{Chang2013,Chang2015}. It can also be achieved in thin exfoliated flakes of MnBi$_2$Te$_4$\cite{Deng2020}.

When a TI presenting the QAHE is in contact with a conventional $s$-wave superconductor, the combination of electron doping from the superconductor to the TI surface and the superconducting proximity effect is expected to lead to a time-reversal-breaking 2D topological superconductivity \cite{Qi2010}. The existence of ferromagnetism in the TI surface makes the situation different from the case of the proximitization of a pristine TI. Due to the TRS breaking, the proximitized TI surface is expected to be accompanied by chiral Majorana edge states \cite{Qi2010}, which are a dispersive 1D edge mode having the Majorana character (i.e. this mode is self-conjugate) and may also be called chiral Majorana fermions. Experimentally, the generation of chiral Majorana fermions is yet to be achieved.

%In quantum field theory, a self-conjugate particle is its own antiparticle, and it is the decisive character of the original Majorana fermions conceived by Ettore Majorana as a model for neutrinos\cite{Wilczek2009}. In this sense, one may say that this 1D edge mode realizes chiral Majorana fermions. Therefore, proximity-induced superconductivity in a QAH insulator is of particular interest in Majorana physics, and there are already experimental works along this line\cite{He2017,Kayyalha2020}. However, the generation of chiral Majorana fermions remains controversial\cite{Kayyalha2020}.

%
\section{Outlook}

The discovery of TIs has caused a paradigm shift in our understanding of how quantum mechanics govern the properties of materials. It has become recognized that the topological nature of the quantum-mechanical wavefunctions realized in a material has sharply observable consequences. The search for new topological invariants in materials and classifications of materials based on new topological classes are an active forefront of the current condensed matter physics. Also, this new viewpoint for materials has triggered the search for similar topological nature in other systems, such as cold atoms, photonic crystals, mechanical metamaterials, qubit arrays, etc. 

The concept of topological superconductors is related to topological insulators, because both materials have a gap at the Fermi energy -- superconducting gap in the former and the band gap in the latter. Because of the existence of the gap, the same bulk-boundary correspondence applies to topological superconductors and they must be accompanied by gapless boundary states, which often (but not always) bear the Majorana character. When a 2D topological superconductor breaks time reversal symmetry and has no spin degeneracy, its edge harbors chiral Majorana fermions and the vortex core accommodates a MZM. The 3D TI material Bi$_2$Se$_3$ becomes a superconductor upon doping of Cu, Sr, or Nb; the resulting 3D superconductivity has been elucidated to be topological and presents peculiar nematic superconductivity \cite{Yonezawa2019}. The Majorana physics realized in topological superconductors is a very interesting topic in the future, and TIs provide a useful platform to address topological superconductivity. 

\section{Further reading}
\begin{flushleft}

$\bullet$ Y. Ando, Topological Insulator Materials, J. Phys. Soc. Jpn. {\bf 82}, 102001 (2013).

$\bullet$ Y. Ando and L. Fu, Topological Crystalline Insulators and Topological Superconductors: From Concepts to Materials, Annu. Rev. Condens. Matter Phys. {\bf 6}, 361 (2015).

$\bullet$ B. A. Bernevig and T. L. Hughes, Topological Insulators and Topological Superconductors. (Princeton University Press, 2013).

$\bullet$ M. Sato and Y. Ando, Topological superconductors: a review, Rep. Prog. Phys. {\bf 80}, 076501 (2017).

$\bullet$ M. Z. Hasan and C. L. Kane, Colloquium: Topological insulators, Rev. Mod. Phys. {\bf 82}, 3045 (2010).

$\bullet$ X.-L. Qi and S.-C. Zhang, Topological insulators and superconductors, Rev. Mod. Phys. {\bf 83}, 1057 (2011).

$\bullet$ Y. Tokura, K. Yasuda, and A. Tsukazaki, Magnetic topological insulators, Nat. Rev. Phys. {\bf 1}, 126 (2019).

\end{flushleft}

\section{Acknowledgments}

The author has been supported by the European Research Council (ERC) under the European Union's Horizon 2020 research and innovation programme (grant agreement No 741121) and by the Deutsche Forschungsgemeinschaft (DFG, German Research Foundation) under CRC 1238 - 277146847 (Subproject A04) as well as under Germany's Excellence Strategy - Cluster of Excellence Matter and Light for Quantum Computing (ML4Q) EXC 2004/1 - 390534769.

%\bibliography{library}

\begin{thebibliography}{67}%
\makeatletter
\providecommand \@ifxundefined [1]{%
 \@ifx{#1\undefined}
}%
\providecommand \@ifnum [1]{%
 \ifnum #1\expandafter \@firstoftwo
 \else \expandafter \@secondoftwo
 \fi
}%
\providecommand \@ifx [1]{%
 \ifx #1\expandafter \@firstoftwo
 \else \expandafter \@secondoftwo
 \fi
}%
\providecommand \natexlab [1]{#1}%
\providecommand \enquote  [1]{``#1''}%
\providecommand \bibnamefont  [1]{#1}%
\providecommand \bibfnamefont [1]{#1}%
\providecommand \citenamefont [1]{#1}%
\providecommand \href@noop [0]{\@secondoftwo}%
\providecommand \href [0]{\begingroup \@sanitize@url \@href}%
\providecommand \@href[1]{\@@startlink{#1}\@@href}%
\providecommand \@@href[1]{\endgroup#1\@@endlink}%
\providecommand \@sanitize@url [0]{\catcode `\\12\catcode `\$12\catcode
  `\&12\catcode `\#12\catcode `\^12\catcode `\_12\catcode `\%12\relax}%
\providecommand \@@startlink[1]{}%
\providecommand \@@endlink[0]{}%
\providecommand \url  [0]{\begingroup\@sanitize@url \@url }%
\providecommand \@url [1]{\endgroup\@href {#1}{\urlprefix }}%
\providecommand \urlprefix  [0]{URL }%
\providecommand \Eprint [0]{\href }%
\providecommand \doibase [0]{http://dx.doi.org/}%
\providecommand \selectlanguage [0]{\@gobble}%
\providecommand \bibinfo  [0]{\@secondoftwo}%
\providecommand \bibfield  [0]{\@secondoftwo}%
\providecommand \translation [1]{[#1]}%
\providecommand \BibitemOpen [0]{}%
\providecommand \bibitemStop [0]{}%
\providecommand \bibitemNoStop [0]{.\EOS\space}%
\providecommand \EOS [0]{\spacefactor3000\relax}%
\providecommand \BibitemShut  [1]{\csname bibitem#1\endcsname}%
\let\auto@bib@innerbib\@empty
%</preamble>
\bibitem [{\citenamefont {Benalcazar}\ \emph {et~al.}(2017)\citenamefont
  {Benalcazar}, \citenamefont {Bernevig},\ and\ \citenamefont
  {Hughes}}]{Benalcazar2017}%
  \BibitemOpen
  \bibfield  {author} {\bibinfo {author} {\bibnamefont {Benalcazar},
  \bibfnamefont {W.~A.}}, \bibinfo {author} {\bibfnamefont {B.~A.}\
  \bibnamefont {Bernevig}}, \ and\ \bibinfo {author} {\bibfnamefont {T.~L.}\
  \bibnamefont {Hughes}}} (\bibinfo {year} {2017}),\ \href {\doibase
  10.1126/science.aah6442} {\bibfield  {journal} {\bibinfo  {journal}
  {Science}\ }\textbf {\bibinfo {volume} {357}}~(\bibinfo {number} {6346}),\
  \bibinfo {pages} {61}}\BibitemShut {NoStop}%
\bibitem [{\citenamefont {Bernevig}\ \emph {et~al.}(2002)\citenamefont
  {Bernevig}, \citenamefont {Chern}, \citenamefont {Hu}, \citenamefont
  {Toumbas},\ and\ \citenamefont {Zhang}}]{Bernevig2002}%
  \BibitemOpen
  \bibfield  {author} {\bibinfo {author} {\bibnamefont {Bernevig},
  \bibfnamefont {B.~A.}}, \bibinfo {author} {\bibfnamefont {C.~H.}\
  \bibnamefont {Chern}}, \bibinfo {author} {\bibfnamefont {J.~P.}\ \bibnamefont
  {Hu}}, \bibinfo {author} {\bibfnamefont {N.}~\bibnamefont {Toumbas}}, \ and\
  \bibinfo {author} {\bibfnamefont {S.~C.}\ \bibnamefont {Zhang}}} (\bibinfo
  {year} {2002}),\ \href {\doibase 10.1006/aphy.2002.6292} {\bibfield
  {journal} {\bibinfo  {journal} {Ann. Phys.}\ }\textbf {\bibinfo {volume}
  {300}}~(\bibinfo {number} {2}),\ \bibinfo {pages} {185}}\BibitemShut
  {NoStop}%
\bibitem [{\citenamefont {Bernevig}\ \emph {et~al.}(2006)\citenamefont
  {Bernevig}, \citenamefont {Hughes},\ and\ \citenamefont
  {Zhang}}]{Bernevig2006}%
  \BibitemOpen
  \bibfield  {author} {\bibinfo {author} {\bibnamefont {Bernevig},
  \bibfnamefont {B.~A.}}, \bibinfo {author} {\bibfnamefont {T.~L.}\
  \bibnamefont {Hughes}}, \ and\ \bibinfo {author} {\bibfnamefont {S.-C.}\
  \bibnamefont {Zhang}}} (\bibinfo {year} {2006}),\ \href {\doibase
  10.1126/science.1133734} {\bibfield  {journal} {\bibinfo  {journal}
  {Science}\ }\textbf {\bibinfo {volume} {314}}~(\bibinfo {number} {5806}),\
  \bibinfo {pages} {1757}}\BibitemShut {NoStop}%
\bibitem [{\citenamefont {Breunig}\ and\ \citenamefont
  {Ando}(2021)}]{Breunig2021}%
  \BibitemOpen
  \bibfield  {author} {\bibinfo {author} {\bibnamefont {Breunig}, \bibfnamefont
  {O.}}, \ and\ \bibinfo {author} {\bibfnamefont {Y.}~\bibnamefont {Ando}}}
  (\bibinfo {year} {2021}),\ \href {\doibase 10.1038/s42254-021-00402-6}
  {\bibfield  {journal} {\bibinfo  {journal} {Nat. Rev. Phys.}\
  }10.1038/s42254-021-00402-6}\BibitemShut {NoStop}%
\bibitem [{\citenamefont {Breunig}\ \emph {et~al.}(2017)\citenamefont
  {Breunig}, \citenamefont {Wang}, \citenamefont {Taskin}, \citenamefont {Lux},
  \citenamefont {Rosch},\ and\ \citenamefont {Ando}}]{Breunig2017}%
  \BibitemOpen
  \bibfield  {author} {\bibinfo {author} {\bibnamefont {Breunig}, \bibfnamefont
  {O.}}, \bibinfo {author} {\bibfnamefont {Z.}~\bibnamefont {Wang}}, \bibinfo
  {author} {\bibfnamefont {A.~A.}\ \bibnamefont {Taskin}}, \bibinfo {author}
  {\bibfnamefont {J.}~\bibnamefont {Lux}}, \bibinfo {author} {\bibfnamefont
  {A.}~\bibnamefont {Rosch}}, \ and\ \bibinfo {author} {\bibfnamefont
  {Y.}~\bibnamefont {Ando}}} (\bibinfo {year} {2017}),\ \href {\doibase
  10.1038/ncomms15545} {\bibfield  {journal} {\bibinfo  {journal} {Nat.
  Commun.}\ }\textbf {\bibinfo {volume} {8}},\ \bibinfo {pages}
  {15545}}\BibitemShut {NoStop}%
\bibitem [{\citenamefont {Bruene}\ \emph {et~al.}(2011)\citenamefont {Bruene},
  \citenamefont {Liu}, \citenamefont {Novik}, \citenamefont {Hankiewicz},
  \citenamefont {Buhmann}, \citenamefont {Chen}, \citenamefont {Qi},
  \citenamefont {Shen}, \citenamefont {Zhang},\ and\ \citenamefont
  {Molenkamp}}]{Bruene2011}%
  \BibitemOpen
  \bibfield  {author} {\bibinfo {author} {\bibnamefont {Bruene}, \bibfnamefont
  {C.}}, \bibinfo {author} {\bibfnamefont {C.~X.}\ \bibnamefont {Liu}},
  \bibinfo {author} {\bibfnamefont {E.~G.}\ \bibnamefont {Novik}}, \bibinfo
  {author} {\bibfnamefont {E.~M.}\ \bibnamefont {Hankiewicz}}, \bibinfo
  {author} {\bibfnamefont {H.}~\bibnamefont {Buhmann}}, \bibinfo {author}
  {\bibfnamefont {Y.~L.}\ \bibnamefont {Chen}}, \bibinfo {author}
  {\bibfnamefont {X.~L.}\ \bibnamefont {Qi}}, \bibinfo {author} {\bibfnamefont
  {Z.~X.}\ \bibnamefont {Shen}}, \bibinfo {author} {\bibfnamefont {S.~C.}\
  \bibnamefont {Zhang}}, \ and\ \bibinfo {author} {\bibfnamefont {L.~W.}\
  \bibnamefont {Molenkamp}}} (\bibinfo {year} {2011}),\ \href {\doibase
  10.1103/PhysRevLett.106.126803} {\bibfield  {journal} {\bibinfo  {journal}
  {Phys. Rev. Lett.}\ }\textbf {\bibinfo {volume} {106}}~(\bibinfo {number}
  {12}),\ \bibinfo {pages} {126803}}\BibitemShut {NoStop}%
\bibitem [{\citenamefont {Bruene}\ \emph {et~al.}(2012)\citenamefont {Bruene},
  \citenamefont {Roth}, \citenamefont {Buhmann}, \citenamefont {Hankiewicz},
  \citenamefont {Molenkamp}, \citenamefont {Maciejko}, \citenamefont {Qi},\
  and\ \citenamefont {Zhang}}]{Bruene2012}%
  \BibitemOpen
  \bibfield  {author} {\bibinfo {author} {\bibnamefont {Bruene}, \bibfnamefont
  {C.}}, \bibinfo {author} {\bibfnamefont {A.}~\bibnamefont {Roth}}, \bibinfo
  {author} {\bibfnamefont {H.}~\bibnamefont {Buhmann}}, \bibinfo {author}
  {\bibfnamefont {E.~M.}\ \bibnamefont {Hankiewicz}}, \bibinfo {author}
  {\bibfnamefont {L.~W.}\ \bibnamefont {Molenkamp}}, \bibinfo {author}
  {\bibfnamefont {J.}~\bibnamefont {Maciejko}}, \bibinfo {author}
  {\bibfnamefont {X.-L.}\ \bibnamefont {Qi}}, \ and\ \bibinfo {author}
  {\bibfnamefont {S.-C.}\ \bibnamefont {Zhang}}} (\bibinfo {year} {2012}),\
  \href {\doibase 10.1038/nphys2322} {\bibfield  {journal} {\bibinfo  {journal}
  {Nature Physics}\ }\textbf {\bibinfo {volume} {8}}~(\bibinfo {number} {6}),\
  \bibinfo {pages} {485}}\BibitemShut {NoStop}%
\bibitem [{\citenamefont {Chang}\ \emph {et~al.}(2013)\citenamefont {Chang},
  \citenamefont {Zhang}, \citenamefont {Feng}, \citenamefont {Shen},
  \citenamefont {Zhang}, \citenamefont {Guo}, \citenamefont {Li}, \citenamefont
  {Ou}, \citenamefont {Wei}, \citenamefont {Wang}, \citenamefont {Ji},
  \citenamefont {Feng}, \citenamefont {Ji}, \citenamefont {Chen}, \citenamefont
  {Jia}, \citenamefont {Dai}, \citenamefont {Fang}, \citenamefont {Zhang},
  \citenamefont {He}, \citenamefont {Wang}, \citenamefont {Lu}, \citenamefont
  {Ma},\ and\ \citenamefont {Xue}}]{Chang2013}%
  \BibitemOpen
  \bibfield  {author} {\bibinfo {author} {\bibnamefont {Chang}, \bibfnamefont
  {C.-Z.}}, \bibinfo {author} {\bibfnamefont {J.}~\bibnamefont {Zhang}},
  \bibinfo {author} {\bibfnamefont {X.}~\bibnamefont {Feng}}, \bibinfo {author}
  {\bibfnamefont {J.}~\bibnamefont {Shen}}, \bibinfo {author} {\bibfnamefont
  {Z.}~\bibnamefont {Zhang}}, \bibinfo {author} {\bibfnamefont
  {M.}~\bibnamefont {Guo}}, \bibinfo {author} {\bibfnamefont {K.}~\bibnamefont
  {Li}}, \bibinfo {author} {\bibfnamefont {Y.}~\bibnamefont {Ou}}, \bibinfo
  {author} {\bibfnamefont {P.}~\bibnamefont {Wei}}, \bibinfo {author}
  {\bibfnamefont {L.-L.}\ \bibnamefont {Wang}}, \bibinfo {author}
  {\bibfnamefont {Z.-Q.}\ \bibnamefont {Ji}}, \bibinfo {author} {\bibfnamefont
  {Y.}~\bibnamefont {Feng}}, \bibinfo {author} {\bibfnamefont {S.}~\bibnamefont
  {Ji}}, \bibinfo {author} {\bibfnamefont {X.}~\bibnamefont {Chen}}, \bibinfo
  {author} {\bibfnamefont {J.}~\bibnamefont {Jia}}, \bibinfo {author}
  {\bibfnamefont {X.}~\bibnamefont {Dai}}, \bibinfo {author} {\bibfnamefont
  {Z.}~\bibnamefont {Fang}}, \bibinfo {author} {\bibfnamefont {S.-C.}\
  \bibnamefont {Zhang}}, \bibinfo {author} {\bibfnamefont {K.}~\bibnamefont
  {He}}, \bibinfo {author} {\bibfnamefont {Y.}~\bibnamefont {Wang}}, \bibinfo
  {author} {\bibfnamefont {L.}~\bibnamefont {Lu}}, \bibinfo {author}
  {\bibfnamefont {X.-C.}\ \bibnamefont {Ma}}, \ and\ \bibinfo {author}
  {\bibfnamefont {Q.-K.}\ \bibnamefont {Xue}}} (\bibinfo {year} {2013}),\ \href
  {\doibase 10.1126/science.1234414} {\bibfield  {journal} {\bibinfo  {journal}
  {Science}\ }\textbf {\bibinfo {volume} {340}}~(\bibinfo {number} {6129}),\
  \bibinfo {pages} {167}}\BibitemShut {NoStop}%
\bibitem [{\citenamefont {Chang}\ \emph {et~al.}(2015)\citenamefont {Chang},
  \citenamefont {Zhao}, \citenamefont {Kim}, \citenamefont {Zhang},
  \citenamefont {Assaf}, \citenamefont {Heiman}, \citenamefont {Zhang},
  \citenamefont {Liu}, \citenamefont {Chan},\ and\ \citenamefont
  {Moodera}}]{Chang2015}%
  \BibitemOpen
  \bibfield  {author} {\bibinfo {author} {\bibnamefont {Chang}, \bibfnamefont
  {C.-Z.}}, \bibinfo {author} {\bibfnamefont {W.}~\bibnamefont {Zhao}},
  \bibinfo {author} {\bibfnamefont {D.~Y.}\ \bibnamefont {Kim}}, \bibinfo
  {author} {\bibfnamefont {H.}~\bibnamefont {Zhang}}, \bibinfo {author}
  {\bibfnamefont {B.~A.}\ \bibnamefont {Assaf}}, \bibinfo {author}
  {\bibfnamefont {D.}~\bibnamefont {Heiman}}, \bibinfo {author} {\bibfnamefont
  {S.-C.}\ \bibnamefont {Zhang}}, \bibinfo {author} {\bibfnamefont
  {C.}~\bibnamefont {Liu}}, \bibinfo {author} {\bibfnamefont {M.~H.~W.}\
  \bibnamefont {Chan}}, \ and\ \bibinfo {author} {\bibfnamefont {J.~S.}\
  \bibnamefont {Moodera}}} (\bibinfo {year} {2015}),\ \href@noop {} {\bibfield
  {journal} {\bibinfo  {journal} {Nat. Mater.}\ }\textbf {\bibinfo {volume}
  {14}}~(\bibinfo {number} {5}),\ \bibinfo {pages} {473}}\BibitemShut {NoStop}%
\bibitem [{\citenamefont {Chen}\ \emph {et~al.}(2009)\citenamefont {Chen},
  \citenamefont {Analytis}, \citenamefont {Chu}, \citenamefont {Liu},
  \citenamefont {Mo}, \citenamefont {Qi}, \citenamefont {Zhang}, \citenamefont
  {Lu}, \citenamefont {Dai}, \citenamefont {Fang}, \citenamefont {Zhang},
  \citenamefont {Fisher}, \citenamefont {Hussain},\ and\ \citenamefont
  {Shen}}]{Chen2009}%
  \BibitemOpen
  \bibfield  {author} {\bibinfo {author} {\bibnamefont {Chen}, \bibfnamefont
  {Y.~L.}}, \bibinfo {author} {\bibfnamefont {J.~G.}\ \bibnamefont {Analytis}},
  \bibinfo {author} {\bibfnamefont {J.~H.}\ \bibnamefont {Chu}}, \bibinfo
  {author} {\bibfnamefont {Z.~K.}\ \bibnamefont {Liu}}, \bibinfo {author}
  {\bibfnamefont {S.~K.}\ \bibnamefont {Mo}}, \bibinfo {author} {\bibfnamefont
  {X.~L.}\ \bibnamefont {Qi}}, \bibinfo {author} {\bibfnamefont {H.~J.}\
  \bibnamefont {Zhang}}, \bibinfo {author} {\bibfnamefont {D.~H.}\ \bibnamefont
  {Lu}}, \bibinfo {author} {\bibfnamefont {X.}~\bibnamefont {Dai}}, \bibinfo
  {author} {\bibfnamefont {Z.}~\bibnamefont {Fang}}, \bibinfo {author}
  {\bibfnamefont {S.~C.}\ \bibnamefont {Zhang}}, \bibinfo {author}
  {\bibfnamefont {I.~R.}\ \bibnamefont {Fisher}}, \bibinfo {author}
  {\bibfnamefont {Z.}~\bibnamefont {Hussain}}, \ and\ \bibinfo {author}
  {\bibfnamefont {Z.~X.}\ \bibnamefont {Shen}}} (\bibinfo {year} {2009}),\
  \href {\doibase 10.1126/science.1173034} {\bibfield  {journal} {\bibinfo
  {journal} {Science}\ }\textbf {\bibinfo {volume} {325}}~(\bibinfo {number}
  {5937}),\ \bibinfo {pages} {178}}\BibitemShut {NoStop}%
\bibitem [{\citenamefont {Cheng}\ \emph {et~al.}(2010)\citenamefont {Cheng},
  \citenamefont {Song}, \citenamefont {Zhang}, \citenamefont {Zhang},
  \citenamefont {Wang}, \citenamefont {Jia}, \citenamefont {Wang},
  \citenamefont {Wang}, \citenamefont {Zhu}, \citenamefont {Chen},
  \citenamefont {Ma}, \citenamefont {He}, \citenamefont {Wang}, \citenamefont
  {Dai}, \citenamefont {Fang}, \citenamefont {Xie}, \citenamefont {Qi},
  \citenamefont {Liu}, \citenamefont {Zhang},\ and\ \citenamefont
  {Xue}}]{Chen2010}%
  \BibitemOpen
  \bibfield  {author} {\bibinfo {author} {\bibnamefont {Cheng}, \bibfnamefont
  {P.}}, \bibinfo {author} {\bibfnamefont {C.}~\bibnamefont {Song}}, \bibinfo
  {author} {\bibfnamefont {T.}~\bibnamefont {Zhang}}, \bibinfo {author}
  {\bibfnamefont {Y.}~\bibnamefont {Zhang}}, \bibinfo {author} {\bibfnamefont
  {Y.}~\bibnamefont {Wang}}, \bibinfo {author} {\bibfnamefont {J.-F.}\
  \bibnamefont {Jia}}, \bibinfo {author} {\bibfnamefont {J.}~\bibnamefont
  {Wang}}, \bibinfo {author} {\bibfnamefont {Y.}~\bibnamefont {Wang}}, \bibinfo
  {author} {\bibfnamefont {B.-F.}\ \bibnamefont {Zhu}}, \bibinfo {author}
  {\bibfnamefont {X.}~\bibnamefont {Chen}}, \bibinfo {author} {\bibfnamefont
  {X.}~\bibnamefont {Ma}}, \bibinfo {author} {\bibfnamefont {K.}~\bibnamefont
  {He}}, \bibinfo {author} {\bibfnamefont {L.}~\bibnamefont {Wang}}, \bibinfo
  {author} {\bibfnamefont {X.}~\bibnamefont {Dai}}, \bibinfo {author}
  {\bibfnamefont {Z.}~\bibnamefont {Fang}}, \bibinfo {author} {\bibfnamefont
  {X.}~\bibnamefont {Xie}}, \bibinfo {author} {\bibfnamefont {X.-L.}\
  \bibnamefont {Qi}}, \bibinfo {author} {\bibfnamefont {C.-X.}\ \bibnamefont
  {Liu}}, \bibinfo {author} {\bibfnamefont {S.-C.}\ \bibnamefont {Zhang}}, \
  and\ \bibinfo {author} {\bibfnamefont {Q.-K.}\ \bibnamefont {Xue}}} (\bibinfo
  {year} {2010}),\ \href {\doibase 10.1103/PhysRevLett.105.076801} {\bibfield
  {journal} {\bibinfo  {journal} {Phys. Rev. Lett.}\ }\textbf {\bibinfo
  {volume} {105}}~(\bibinfo {number} {7}),\ \bibinfo {pages}
  {076801}}\BibitemShut {NoStop}%
\bibitem [{\citenamefont {Cook}\ and\ \citenamefont {Franz}(2011)}]{Cook2011}%
  \BibitemOpen
  \bibfield  {author} {\bibinfo {author} {\bibnamefont {Cook}, \bibfnamefont
  {A.}}, \ and\ \bibinfo {author} {\bibfnamefont {M.}~\bibnamefont {Franz}}}
  (\bibinfo {year} {2011}),\ \href {\doibase 10.1103/PhysRevB.84.201105}
  {\bibfield  {journal} {\bibinfo  {journal} {Phys. Rev. B}\ }\textbf {\bibinfo
  {volume} {84}}~(\bibinfo {number} {20}),\ \bibinfo {pages}
  {201105}}\BibitemShut {NoStop}%
\bibitem [{\citenamefont {Deng}\ \emph {et~al.}(2020)\citenamefont {Deng},
  \citenamefont {Yu}, \citenamefont {Shi}, \citenamefont {Guo}, \citenamefont
  {Xu}, \citenamefont {Wang}, \citenamefont {Chen},\ and\ \citenamefont
  {Zhang}}]{Deng2020}%
  \BibitemOpen
  \bibfield  {author} {\bibinfo {author} {\bibnamefont {Deng}, \bibfnamefont
  {Y.}}, \bibinfo {author} {\bibfnamefont {Y.}~\bibnamefont {Yu}}, \bibinfo
  {author} {\bibfnamefont {M.~Z.}\ \bibnamefont {Shi}}, \bibinfo {author}
  {\bibfnamefont {Z.}~\bibnamefont {Guo}}, \bibinfo {author} {\bibfnamefont
  {Z.}~\bibnamefont {Xu}}, \bibinfo {author} {\bibfnamefont {J.}~\bibnamefont
  {Wang}}, \bibinfo {author} {\bibfnamefont {X.~H.}\ \bibnamefont {Chen}}, \
  and\ \bibinfo {author} {\bibfnamefont {Y.}~\bibnamefont {Zhang}}} (\bibinfo
  {year} {2020}),\ \href {\doibase 10.1126/science.aax8156} {\bibfield
  {journal} {\bibinfo  {journal} {Science}\ }\textbf {\bibinfo {volume}
  {367}}~(\bibinfo {number} {6480}),\ \bibinfo {pages} {895}}\BibitemShut
  {NoStop}%
\bibitem [{\citenamefont {Dzero}\ \emph {et~al.}(2010)\citenamefont {Dzero},
  \citenamefont {Sun}, \citenamefont {Galitski},\ and\ \citenamefont
  {Coleman}}]{Dzero2010}%
  \BibitemOpen
  \bibfield  {author} {\bibinfo {author} {\bibnamefont {Dzero}, \bibfnamefont
  {M.}}, \bibinfo {author} {\bibfnamefont {K.}~\bibnamefont {Sun}}, \bibinfo
  {author} {\bibfnamefont {V.}~\bibnamefont {Galitski}}, \ and\ \bibinfo
  {author} {\bibfnamefont {P.}~\bibnamefont {Coleman}}} (\bibinfo {year}
  {2010}),\ \href {\doibase 10.1103/PhysRevLett.104.106408} {\bibfield
  {journal} {\bibinfo  {journal} {Phys. Rev. Lett.}\ }\textbf {\bibinfo
  {volume} {104}}~(\bibinfo {number} {10}),\ \bibinfo {pages}
  {106408}}\BibitemShut {NoStop}%
\bibitem [{\citenamefont {Dziawa}\ \emph {et~al.}(2012)\citenamefont {Dziawa},
  \citenamefont {Kowalski}, \citenamefont {Dybko}, \citenamefont {Buczko},
  \citenamefont {Szczerbakow}, \citenamefont {Szot}, \citenamefont
  {Lusakowska}, \citenamefont {Balasubramanian}, \citenamefont {Wojek},
  \citenamefont {Berntsen}, \citenamefont {Tjernberg},\ and\ \citenamefont
  {Story}}]{Dziawa2012}%
  \BibitemOpen
  \bibfield  {author} {\bibinfo {author} {\bibnamefont {Dziawa}, \bibfnamefont
  {P.}}, \bibinfo {author} {\bibfnamefont {B.~J.}\ \bibnamefont {Kowalski}},
  \bibinfo {author} {\bibfnamefont {K.}~\bibnamefont {Dybko}}, \bibinfo
  {author} {\bibfnamefont {R.}~\bibnamefont {Buczko}}, \bibinfo {author}
  {\bibfnamefont {A.}~\bibnamefont {Szczerbakow}}, \bibinfo {author}
  {\bibfnamefont {M.}~\bibnamefont {Szot}}, \bibinfo {author} {\bibfnamefont
  {E.}~\bibnamefont {Lusakowska}}, \bibinfo {author} {\bibfnamefont
  {T.}~\bibnamefont {Balasubramanian}}, \bibinfo {author} {\bibfnamefont
  {B.~M.}\ \bibnamefont {Wojek}}, \bibinfo {author} {\bibfnamefont {M.~H.}\
  \bibnamefont {Berntsen}}, \bibinfo {author} {\bibfnamefont {O.}~\bibnamefont
  {Tjernberg}}, \ and\ \bibinfo {author} {\bibfnamefont {T.}~\bibnamefont
  {Story}}} (\bibinfo {year} {2012}),\ \href {\doibase 10.1038/nmat3449}
  {\bibfield  {journal} {\bibinfo  {journal} {Nat. Mater.}\ }\textbf {\bibinfo
  {volume} {11}}~(\bibinfo {number} {12}),\ \bibinfo {pages}
  {1023}}\BibitemShut {NoStop}%
\bibitem [{\citenamefont {Fu}(2011)}]{Fu2011}%
  \BibitemOpen
  \bibfield  {author} {\bibinfo {author} {\bibnamefont {Fu}, \bibfnamefont
  {L.}}} (\bibinfo {year} {2011}),\ \href@noop {} {\bibfield  {journal}
  {\bibinfo  {journal} {Phys. Rev. Lett.}\ }\textbf {\bibinfo {volume} {106}},\
  \bibinfo {pages} {106802}}\BibitemShut {NoStop}%
\bibitem [{\citenamefont {Fu}\ and\ \citenamefont {Kane}(2007)}]{Fu2007b}%
  \BibitemOpen
  \bibfield  {author} {\bibinfo {author} {\bibnamefont {Fu}, \bibfnamefont
  {L.}}, \ and\ \bibinfo {author} {\bibfnamefont {C.~L.}\ \bibnamefont {Kane}}}
  (\bibinfo {year} {2007}),\ \href@noop {} {\bibfield  {journal} {\bibinfo
  {journal} {Phys. Rev. B}\ }\textbf {\bibinfo {volume} {76}},\ \bibinfo
  {pages} {045302}}\BibitemShut {NoStop}%
\bibitem [{\citenamefont {Fu}\ and\ \citenamefont {Kane}(2008)}]{Fu2008}%
  \BibitemOpen
  \bibfield  {author} {\bibinfo {author} {\bibnamefont {Fu}, \bibfnamefont
  {L.}}, \ and\ \bibinfo {author} {\bibfnamefont {C.~L.}\ \bibnamefont {Kane}}}
  (\bibinfo {year} {2008}),\ \href@noop {} {\bibfield  {journal} {\bibinfo
  {journal} {Phys. Rev. Lett.}\ }\textbf {\bibinfo {volume} {100}}~(\bibinfo
  {number} {9}),\ \bibinfo {pages} {096407}}\BibitemShut {NoStop}%
\bibitem [{\citenamefont {Fu}\ \emph {et~al.}(2007)\citenamefont {Fu},
  \citenamefont {Kane},\ and\ \citenamefont {Mele}}]{Fu2007}%
  \BibitemOpen
  \bibfield  {author} {\bibinfo {author} {\bibnamefont {Fu}, \bibfnamefont
  {L.}}, \bibinfo {author} {\bibfnamefont {C.~L.}\ \bibnamefont {Kane}}, \ and\
  \bibinfo {author} {\bibfnamefont {E.~J.}\ \bibnamefont {Mele}}} (\bibinfo
  {year} {2007}),\ \href@noop {} {\bibfield  {journal} {\bibinfo  {journal}
  {Phys. Rev. Lett.}\ }\textbf {\bibinfo {volume} {98}}~(\bibinfo {number}
  {10}),\ \bibinfo {pages} {106803}}\BibitemShut {NoStop}%
\bibitem [{\citenamefont {Hsieh}\ \emph {et~al.}(2008)\citenamefont {Hsieh},
  \citenamefont {Qian}, \citenamefont {Wray}, \citenamefont {Xia},
  \citenamefont {Hor}, \citenamefont {Cava},\ and\ \citenamefont
  {Hasan}}]{Hsieh2008}%
  \BibitemOpen
  \bibfield  {author} {\bibinfo {author} {\bibnamefont {Hsieh}, \bibfnamefont
  {D.}}, \bibinfo {author} {\bibfnamefont {D.}~\bibnamefont {Qian}}, \bibinfo
  {author} {\bibfnamefont {L.}~\bibnamefont {Wray}}, \bibinfo {author}
  {\bibfnamefont {Y.}~\bibnamefont {Xia}}, \bibinfo {author} {\bibfnamefont
  {Y.~S.}\ \bibnamefont {Hor}}, \bibinfo {author} {\bibfnamefont {R.~J.}\
  \bibnamefont {Cava}}, \ and\ \bibinfo {author} {\bibfnamefont {M.~Z.}\
  \bibnamefont {Hasan}}} (\bibinfo {year} {2008}),\ \href {\doibase
  10.1038/nature06843} {\bibfield  {journal} {\bibinfo  {journal} {Nature}\
  }\textbf {\bibinfo {volume} {452}}~(\bibinfo {number} {7190}),\ \bibinfo
  {pages} {970}}\BibitemShut {NoStop}%
\bibitem [{\citenamefont {Hsieh}\ \emph {et~al.}(2009)\citenamefont {Hsieh},
  \citenamefont {Xia}, \citenamefont {Wray}, \citenamefont {Qian},
  \citenamefont {Pal}, \citenamefont {Dil}, \citenamefont {Osterwalder},
  \citenamefont {Meier}, \citenamefont {Bihlmayer}, \citenamefont {Kane},
  \citenamefont {Hor}, \citenamefont {Cava},\ and\ \citenamefont
  {Hasan}}]{Hsieh2009}%
  \BibitemOpen
  \bibfield  {author} {\bibinfo {author} {\bibnamefont {Hsieh}, \bibfnamefont
  {D.}}, \bibinfo {author} {\bibfnamefont {Y.}~\bibnamefont {Xia}}, \bibinfo
  {author} {\bibfnamefont {L.}~\bibnamefont {Wray}}, \bibinfo {author}
  {\bibfnamefont {D.}~\bibnamefont {Qian}}, \bibinfo {author} {\bibfnamefont
  {A.}~\bibnamefont {Pal}}, \bibinfo {author} {\bibfnamefont {J.~H.}\
  \bibnamefont {Dil}}, \bibinfo {author} {\bibfnamefont {J.}~\bibnamefont
  {Osterwalder}}, \bibinfo {author} {\bibfnamefont {F.}~\bibnamefont {Meier}},
  \bibinfo {author} {\bibfnamefont {G.}~\bibnamefont {Bihlmayer}}, \bibinfo
  {author} {\bibfnamefont {C.~L.}\ \bibnamefont {Kane}}, \bibinfo {author}
  {\bibfnamefont {Y.~S.}\ \bibnamefont {Hor}}, \bibinfo {author} {\bibfnamefont
  {R.~J.}\ \bibnamefont {Cava}}, \ and\ \bibinfo {author} {\bibfnamefont
  {M.~Z.}\ \bibnamefont {Hasan}}} (\bibinfo {year} {2009}),\ \href {\doibase
  10.1126/science.1167733} {\bibfield  {journal} {\bibinfo  {journal}
  {Science}\ }\textbf {\bibinfo {volume} {323}}~(\bibinfo {number} {5916}),\
  \bibinfo {pages} {919}}\BibitemShut {NoStop}%
\bibitem [{\citenamefont {Hsieh}\ \emph {et~al.}(2012)\citenamefont {Hsieh},
  \citenamefont {Lin}, \citenamefont {Liu}, \citenamefont {Duan}, \citenamefont
  {Bansil},\ and\ \citenamefont {Fu}}]{Hsieh2012}%
  \BibitemOpen
  \bibfield  {author} {\bibinfo {author} {\bibnamefont {Hsieh}, \bibfnamefont
  {T.~H.}}, \bibinfo {author} {\bibfnamefont {H.}~\bibnamefont {Lin}}, \bibinfo
  {author} {\bibfnamefont {J.}~\bibnamefont {Liu}}, \bibinfo {author}
  {\bibfnamefont {W.}~\bibnamefont {Duan}}, \bibinfo {author} {\bibfnamefont
  {A.}~\bibnamefont {Bansil}}, \ and\ \bibinfo {author} {\bibfnamefont
  {L.}~\bibnamefont {Fu}}} (\bibinfo {year} {2012}),\ \href@noop {} {\bibfield
  {journal} {\bibinfo  {journal} {Nat. Commun.}\ }\textbf {\bibinfo {volume}
  {3}},\ \bibinfo {pages} {982}}\BibitemShut {NoStop}%
\bibitem [{\citenamefont {Kane}\ and\ \citenamefont
  {Mele}(2005{\natexlab{a}})}]{Kane2005a}%
  \BibitemOpen
  \bibfield  {author} {\bibinfo {author} {\bibnamefont {Kane}, \bibfnamefont
  {C.~L.}}, \ and\ \bibinfo {author} {\bibfnamefont {E.~J.}\ \bibnamefont
  {Mele}}} (\bibinfo {year} {2005}{\natexlab{a}}),\ \href@noop {} {\bibfield
  {journal} {\bibinfo  {journal} {Phys. Rev. Lett.}\ }\textbf {\bibinfo
  {volume} {95}}~(\bibinfo {number} {22}),\ \bibinfo {pages}
  {226801}}\BibitemShut {NoStop}%
\bibitem [{\citenamefont {Kane}\ and\ \citenamefont
  {Mele}(2005{\natexlab{b}})}]{Kane2005b}%
  \BibitemOpen
  \bibfield  {author} {\bibinfo {author} {\bibnamefont {Kane}, \bibfnamefont
  {C.~L.}}, \ and\ \bibinfo {author} {\bibfnamefont {E.~J.}\ \bibnamefont
  {Mele}}} (\bibinfo {year} {2005}{\natexlab{b}}),\ \href@noop {} {\bibfield
  {journal} {\bibinfo  {journal} {Phys. Rev. Lett.}\ }\textbf {\bibinfo
  {volume} {95}}~(\bibinfo {number} {14}),\ \bibinfo {pages}
  {146802}}\BibitemShut {NoStop}%
\bibitem [{\citenamefont {Kato}\ \emph {et~al.}(2004)\citenamefont {Kato},
  \citenamefont {Myers}, \citenamefont {Gossard},\ and\ \citenamefont
  {Awschalom}}]{Kato2004}%
  \BibitemOpen
  \bibfield  {author} {\bibinfo {author} {\bibnamefont {Kato}, \bibfnamefont
  {Y.~K.}}, \bibinfo {author} {\bibfnamefont {R.~C.}\ \bibnamefont {Myers}},
  \bibinfo {author} {\bibfnamefont {A.~C.}\ \bibnamefont {Gossard}}, \ and\
  \bibinfo {author} {\bibfnamefont {D.~D.}\ \bibnamefont {Awschalom}}}
  (\bibinfo {year} {2004}),\ \href {\doibase 10.1126/science.1105514}
  {\bibfield  {journal} {\bibinfo  {journal} {Science}\ }\textbf {\bibinfo
  {volume} {306}}~(\bibinfo {number} {5703}),\ \bibinfo {pages}
  {1910}}\BibitemShut {NoStop}%
\bibitem [{\citenamefont {von Klitzing}\ and\ \citenamefont
  {Landwehr}(1971)}]{Klitzing1971}%
  \BibitemOpen
  \bibfield  {author} {\bibinfo {author} {\bibnamefont {von Klitzing},
  \bibfnamefont {K.}}, \ and\ \bibinfo {author} {\bibfnamefont
  {G.}~\bibnamefont {Landwehr}}} (\bibinfo {year} {1971}),\ \href {\doibase
  https://doi.org/10.1016/0038-1098(71)90630-2} {\bibfield  {journal} {\bibinfo
   {journal} {Solid State Commun.}\ }\textbf {\bibinfo {volume} {9}}~(\bibinfo
  {number} {24}),\ \bibinfo {pages} {2201}}\BibitemShut {NoStop}%
\bibitem [{\citenamefont {Klitzing}\ \emph {et~al.}(1980)\citenamefont
  {Klitzing}, \citenamefont {Dorda},\ and\ \citenamefont
  {Pepper}}]{Klitzing1980}%
  \BibitemOpen
  \bibfield  {author} {\bibinfo {author} {\bibnamefont {Klitzing},
  \bibfnamefont {K.~v.}}, \bibinfo {author} {\bibfnamefont {G.}~\bibnamefont
  {Dorda}}, \ and\ \bibinfo {author} {\bibfnamefont {M.}~\bibnamefont
  {Pepper}}} (\bibinfo {year} {1980}),\ \href {\doibase
  10.1103/PhysRevLett.45.494} {\bibfield  {journal} {\bibinfo  {journal} {Phys.
  Rev. Lett.}\ }\textbf {\bibinfo {volume} {45}}~(\bibinfo {number} {6}),\
  \bibinfo {pages} {494}}\BibitemShut {NoStop}%
\bibitem [{\citenamefont {Knez}\ \emph {et~al.}(2011)\citenamefont {Knez},
  \citenamefont {Du},\ and\ \citenamefont {Sullivan}}]{Knez2011}%
  \BibitemOpen
  \bibfield  {author} {\bibinfo {author} {\bibnamefont {Knez}, \bibfnamefont
  {I.}}, \bibinfo {author} {\bibfnamefont {R.-R.}\ \bibnamefont {Du}}, \ and\
  \bibinfo {author} {\bibfnamefont {G.}~\bibnamefont {Sullivan}}} (\bibinfo
  {year} {2011}),\ \href {\doibase 10.1103/PhysRevLett.107.136603} {\bibfield
  {journal} {\bibinfo  {journal} {Phys. Rev. Lett.}\ }\textbf {\bibinfo
  {volume} {107}}~(\bibinfo {number} {13}),\ \bibinfo {pages}
  {136603}}\BibitemShut {NoStop}%
\bibitem [{\citenamefont {Kohmoto}(1985)}]{Kohmoto1985}%
  \BibitemOpen
  \bibfield  {author} {\bibinfo {author} {\bibnamefont {Kohmoto}, \bibfnamefont
  {M.}}} (\bibinfo {year} {1985}),\ \href@noop {} {\bibfield  {journal}
  {\bibinfo  {journal} {Ann. Phys.}\ }\textbf {\bibinfo {volume} {160}},\
  \bibinfo {pages} {343}}\BibitemShut {NoStop}%
\bibitem [{\citenamefont {Konig}\ \emph {et~al.}(2007)\citenamefont {Konig},
  \citenamefont {Wiedmann}, \citenamefont {Brune}, \citenamefont {Roth},
  \citenamefont {Buhmann}, \citenamefont {Molenkamp}, \citenamefont {Qi},\ and\
  \citenamefont {Zhang}}]{Konig2007}%
  \BibitemOpen
  \bibfield  {author} {\bibinfo {author} {\bibnamefont {Konig}, \bibfnamefont
  {M.}}, \bibinfo {author} {\bibfnamefont {S.}~\bibnamefont {Wiedmann}},
  \bibinfo {author} {\bibfnamefont {C.}~\bibnamefont {Brune}}, \bibinfo
  {author} {\bibfnamefont {A.}~\bibnamefont {Roth}}, \bibinfo {author}
  {\bibfnamefont {H.}~\bibnamefont {Buhmann}}, \bibinfo {author} {\bibfnamefont
  {L.~W.}\ \bibnamefont {Molenkamp}}, \bibinfo {author} {\bibfnamefont {X.~L.}\
  \bibnamefont {Qi}}, \ and\ \bibinfo {author} {\bibfnamefont {S.~C.}\
  \bibnamefont {Zhang}}} (\bibinfo {year} {2007}),\ \href {\doibase
  10.1126/science.1148047} {\bibfield  {journal} {\bibinfo  {journal}
  {Science}\ }\textbf {\bibinfo {volume} {318}}~(\bibinfo {number} {5851}),\
  \bibinfo {pages} {766}}\BibitemShut {NoStop}%
\bibitem [{\citenamefont {Legg}\ \emph {et~al.}(2021)\citenamefont {Legg},
  \citenamefont {Loss},\ and\ \citenamefont {Klinovaja}}]{Legg2021}%
  \BibitemOpen
  \bibfield  {author} {\bibinfo {author} {\bibnamefont {Legg}, \bibfnamefont
  {H.~F.}}, \bibinfo {author} {\bibfnamefont {D.}~\bibnamefont {Loss}}, \ and\
  \bibinfo {author} {\bibfnamefont {J.}~\bibnamefont {Klinovaja}}} (\bibinfo
  {year} {2021}),\ \href {\doibase 10.1103/PhysRevB.104.165405} {\bibfield
  {journal} {\bibinfo  {journal} {Phys. Rev. B}\ }\textbf {\bibinfo {volume}
  {104}}~(\bibinfo {number} {16}),\ \bibinfo {pages} {165405}}\BibitemShut
  {NoStop}%
\bibitem [{\citenamefont {Liu}\ \emph {et~al.}(2008)\citenamefont {Liu},
  \citenamefont {Hughes}, \citenamefont {Qi}, \citenamefont {Wang},\ and\
  \citenamefont {Zhang}}]{Liu2008}%
  \BibitemOpen
  \bibfield  {author} {\bibinfo {author} {\bibnamefont {Liu}, \bibfnamefont
  {C.}}, \bibinfo {author} {\bibfnamefont {T.~L.}\ \bibnamefont {Hughes}},
  \bibinfo {author} {\bibfnamefont {X.-L.}\ \bibnamefont {Qi}}, \bibinfo
  {author} {\bibfnamefont {K.}~\bibnamefont {Wang}}, \ and\ \bibinfo {author}
  {\bibfnamefont {S.-C.}\ \bibnamefont {Zhang}}} (\bibinfo {year} {2008}),\
  \href {\doibase 10.1103/PhysRevLett.100.236601} {\bibfield  {journal}
  {\bibinfo  {journal} {Phys. Rev. Lett.}\ }\textbf {\bibinfo {volume}
  {100}}~(\bibinfo {number} {23}),\ \bibinfo {pages} {236601}}\BibitemShut
  {NoStop}%
\bibitem [{\citenamefont {Moore}\ and\ \citenamefont
  {Balents}(2007)}]{Moore2007}%
  \BibitemOpen
  \bibfield  {author} {\bibinfo {author} {\bibnamefont {Moore}, \bibfnamefont
  {J.~E.}}, \ and\ \bibinfo {author} {\bibfnamefont {L.}~\bibnamefont
  {Balents}}} (\bibinfo {year} {2007}),\ \href@noop {} {\bibfield  {journal}
  {\bibinfo  {journal} {Phys. Rev. B}\ }\textbf {\bibinfo {volume}
  {75}}~(\bibinfo {number} {12}),\ \bibinfo {pages} {121306}}\BibitemShut
  {NoStop}%
\bibitem [{\citenamefont {Murakami}\ \emph {et~al.}(2004)\citenamefont
  {Murakami}, \citenamefont {Nagaosa},\ and\ \citenamefont
  {Zhang}}]{Murakami2004}%
  \BibitemOpen
  \bibfield  {author} {\bibinfo {author} {\bibnamefont {Murakami},
  \bibfnamefont {S.}}, \bibinfo {author} {\bibfnamefont {N.}~\bibnamefont
  {Nagaosa}}, \ and\ \bibinfo {author} {\bibfnamefont {S.~C.}\ \bibnamefont
  {Zhang}}} (\bibinfo {year} {2004}),\ \href {\doibase
  10.1103/PhysRevLett.93.156804} {\bibfield  {journal} {\bibinfo  {journal}
  {Phys. Rev. Lett.}\ }\textbf {\bibinfo {volume} {93}}~(\bibinfo {number}
  {15}),\ \bibinfo {pages} {156804}}\BibitemShut {NoStop}%
\bibitem [{\citenamefont {Nagaosa}\ \emph {et~al.}(2010)\citenamefont
  {Nagaosa}, \citenamefont {Sinova}, \citenamefont {Onoda}, \citenamefont
  {MacDonald},\ and\ \citenamefont {Ong}}]{Nagaosa2010}%
  \BibitemOpen
  \bibfield  {author} {\bibinfo {author} {\bibnamefont {Nagaosa}, \bibfnamefont
  {N.}}, \bibinfo {author} {\bibfnamefont {J.}~\bibnamefont {Sinova}}, \bibinfo
  {author} {\bibfnamefont {S.}~\bibnamefont {Onoda}}, \bibinfo {author}
  {\bibfnamefont {A.~H.}\ \bibnamefont {MacDonald}}, \ and\ \bibinfo {author}
  {\bibfnamefont {N.~P.}\ \bibnamefont {Ong}}} (\bibinfo {year} {2010}),\ \href
  {\doibase 10.1103/RevModPhys.82.1539} {\bibfield  {journal} {\bibinfo
  {journal} {Rev. Mod. Phy.}\ }\textbf {\bibinfo {volume} {82}}~(\bibinfo
  {number} {2}),\ \bibinfo {pages} {1539}}\BibitemShut {NoStop}%
\bibitem [{\citenamefont {Nayak}\ \emph {et~al.}(2008)\citenamefont {Nayak},
  \citenamefont {Simon}, \citenamefont {Stern}, \citenamefont {Freedman},\ and\
  \citenamefont {Das~Sarma}}]{Nayak2008}%
  \BibitemOpen
  \bibfield  {author} {\bibinfo {author} {\bibnamefont {Nayak}, \bibfnamefont
  {C.}}, \bibinfo {author} {\bibfnamefont {S.~H.}\ \bibnamefont {Simon}},
  \bibinfo {author} {\bibfnamefont {A.}~\bibnamefont {Stern}}, \bibinfo
  {author} {\bibfnamefont {M.}~\bibnamefont {Freedman}}, \ and\ \bibinfo
  {author} {\bibfnamefont {S.}~\bibnamefont {Das~Sarma}}} (\bibinfo {year}
  {2008}),\ \href@noop {} {\bibfield  {journal} {\bibinfo  {journal} {Rev. Mod.
  Phys.}\ }\textbf {\bibinfo {volume} {80}},\ \bibinfo {pages}
  {1083}}\BibitemShut {NoStop}%
\bibitem [{\citenamefont {Nishide}\ \emph {et~al.}(2010)\citenamefont
  {Nishide}, \citenamefont {Taskin}, \citenamefont {Takeichi}, \citenamefont
  {Okuda}, \citenamefont {Kakizaki}, \citenamefont {Hirahara}, \citenamefont
  {Nakatsuji}, \citenamefont {Komori}, \citenamefont {Ando},\ and\
  \citenamefont {Matsuda}}]{Nishide2010}%
  \BibitemOpen
  \bibfield  {author} {\bibinfo {author} {\bibnamefont {Nishide}, \bibfnamefont
  {A.}}, \bibinfo {author} {\bibfnamefont {A.~A.}\ \bibnamefont {Taskin}},
  \bibinfo {author} {\bibfnamefont {Y.}~\bibnamefont {Takeichi}}, \bibinfo
  {author} {\bibfnamefont {T.}~\bibnamefont {Okuda}}, \bibinfo {author}
  {\bibfnamefont {A.}~\bibnamefont {Kakizaki}}, \bibinfo {author}
  {\bibfnamefont {T.}~\bibnamefont {Hirahara}}, \bibinfo {author}
  {\bibfnamefont {K.}~\bibnamefont {Nakatsuji}}, \bibinfo {author}
  {\bibfnamefont {F.}~\bibnamefont {Komori}}, \bibinfo {author} {\bibfnamefont
  {Y.}~\bibnamefont {Ando}}, \ and\ \bibinfo {author} {\bibfnamefont
  {I.}~\bibnamefont {Matsuda}}} (\bibinfo {year} {2010}),\ \href {\doibase
  10.1103/PhysRevB.81.041309} {\bibfield  {journal} {\bibinfo  {journal} {Phys.
  Rev. B}\ }\textbf {\bibinfo {volume} {81}}~(\bibinfo {number} {4}),\ \bibinfo
  {pages} {041309}}\BibitemShut {NoStop}%
\bibitem [{\citenamefont {Onoda}\ and\ \citenamefont
  {Nagaosa}(2005)}]{Onoda2005}%
  \BibitemOpen
  \bibfield  {author} {\bibinfo {author} {\bibnamefont {Onoda}, \bibfnamefont
  {M.}}, \ and\ \bibinfo {author} {\bibfnamefont {N.}~\bibnamefont {Nagaosa}}}
  (\bibinfo {year} {2005}),\ \href {\doibase 10.1103/PhysRevLett.95.106601}
  {\bibfield  {journal} {\bibinfo  {journal} {Phys. Rev. Lett.}\ }\textbf
  {\bibinfo {volume} {95}}~(\bibinfo {number} {10}),\ \bibinfo {pages}
  {106601}}\BibitemShut {NoStop}%
\bibitem [{\citenamefont {Peng}\ \emph {et~al.}(2010)\citenamefont {Peng},
  \citenamefont {Lai}, \citenamefont {Kong}, \citenamefont {Meister},
  \citenamefont {Chen}, \citenamefont {Qi}, \citenamefont {Zhang},
  \citenamefont {Shen},\ and\ \citenamefont {Cui}}]{Peng2010}%
  \BibitemOpen
  \bibfield  {author} {\bibinfo {author} {\bibnamefont {Peng}, \bibfnamefont
  {H.}}, \bibinfo {author} {\bibfnamefont {K.}~\bibnamefont {Lai}}, \bibinfo
  {author} {\bibfnamefont {D.}~\bibnamefont {Kong}}, \bibinfo {author}
  {\bibfnamefont {S.}~\bibnamefont {Meister}}, \bibinfo {author} {\bibfnamefont
  {Y.}~\bibnamefont {Chen}}, \bibinfo {author} {\bibfnamefont {X.-L.}\
  \bibnamefont {Qi}}, \bibinfo {author} {\bibfnamefont {S.-C.}\ \bibnamefont
  {Zhang}}, \bibinfo {author} {\bibfnamefont {Z.-X.}\ \bibnamefont {Shen}}, \
  and\ \bibinfo {author} {\bibfnamefont {Y.}~\bibnamefont {Cui}}} (\bibinfo
  {year} {2010}),\ \href {\doibase 10.1038/nmat2609} {\bibfield  {journal}
  {\bibinfo  {journal} {Nat. Mater.}\ }\textbf {\bibinfo {volume}
  {9}}~(\bibinfo {number} {3}),\ \bibinfo {pages} {225}}\BibitemShut {NoStop}%
\bibitem [{\citenamefont {Qi}\ \emph {et~al.}(2008)\citenamefont {Qi},
  \citenamefont {Hughes},\ and\ \citenamefont {Zhang}}]{Qi2008}%
  \BibitemOpen
  \bibfield  {author} {\bibinfo {author} {\bibnamefont {Qi}, \bibfnamefont
  {X.-L.}}, \bibinfo {author} {\bibfnamefont {T.~L.}\ \bibnamefont {Hughes}}, \
  and\ \bibinfo {author} {\bibfnamefont {S.-C.}\ \bibnamefont {Zhang}}}
  (\bibinfo {year} {2008}),\ \href@noop {} {\bibfield  {journal} {\bibinfo
  {journal} {Phys. Rev. B}\ }\textbf {\bibinfo {volume} {78}},\ \bibinfo
  {pages} {195424}}\BibitemShut {NoStop}%
\bibitem [{\citenamefont {Qi}\ \emph {et~al.}(2010)\citenamefont {Qi},
  \citenamefont {Hughes},\ and\ \citenamefont {Zhang}}]{Qi2010}%
  \BibitemOpen
  \bibfield  {author} {\bibinfo {author} {\bibnamefont {Qi}, \bibfnamefont
  {X.-L.}}, \bibinfo {author} {\bibfnamefont {T.~L.}\ \bibnamefont {Hughes}}, \
  and\ \bibinfo {author} {\bibfnamefont {S.-C.}\ \bibnamefont {Zhang}}}
  (\bibinfo {year} {2010}),\ \href@noop {} {\bibfield  {journal} {\bibinfo
  {journal} {Phys. Rev. B}\ }\textbf {\bibinfo {volume} {82}}~(\bibinfo
  {number} {18}),\ \bibinfo {pages} {184516}}\BibitemShut {NoStop}%
\bibitem [{\citenamefont {Qian}\ \emph {et~al.}(2014)\citenamefont {Qian},
  \citenamefont {Liu}, \citenamefont {Fu},\ and\ \citenamefont
  {Li}}]{Qian2014}%
  \BibitemOpen
  \bibfield  {author} {\bibinfo {author} {\bibnamefont {Qian}, \bibfnamefont
  {X.}}, \bibinfo {author} {\bibfnamefont {J.}~\bibnamefont {Liu}}, \bibinfo
  {author} {\bibfnamefont {L.}~\bibnamefont {Fu}}, \ and\ \bibinfo {author}
  {\bibfnamefont {J.}~\bibnamefont {Li}}} (\bibinfo {year} {2014}),\ \href
  {\doibase 10.1126/science.1256815} {\bibfield  {journal} {\bibinfo  {journal}
  {Science}\ }\textbf {\bibinfo {volume} {346}}~(\bibinfo {number} {6215}),\
  \bibinfo {pages} {1344}}\BibitemShut {NoStop}%
\bibitem [{\citenamefont {Ren}\ \emph {et~al.}(2010)\citenamefont {Ren},
  \citenamefont {Taskin}, \citenamefont {Sasaki}, \citenamefont {Segawa},\ and\
  \citenamefont {Ando}}]{Ren2010}%
  \BibitemOpen
  \bibfield  {author} {\bibinfo {author} {\bibnamefont {Ren}, \bibfnamefont
  {Z.}}, \bibinfo {author} {\bibfnamefont {A.~A.}\ \bibnamefont {Taskin}},
  \bibinfo {author} {\bibfnamefont {S.}~\bibnamefont {Sasaki}}, \bibinfo
  {author} {\bibfnamefont {K.}~\bibnamefont {Segawa}}, \ and\ \bibinfo {author}
  {\bibfnamefont {Y.}~\bibnamefont {Ando}}} (\bibinfo {year} {2010}),\ \href
  {\doibase 10.1103/PhysRevB.82.241306} {\bibfield  {journal} {\bibinfo
  {journal} {Phys. Rev. B}\ }\textbf {\bibinfo {volume} {82}}~(\bibinfo
  {number} {24}),\ \bibinfo {pages} {241306}}\BibitemShut {NoStop}%
\bibitem [{\citenamefont {Ren}\ \emph {et~al.}(2011)\citenamefont {Ren},
  \citenamefont {Taskin}, \citenamefont {Sasaki}, \citenamefont {Segawa},\ and\
  \citenamefont {Ando}}]{Ren2011}%
  \BibitemOpen
  \bibfield  {author} {\bibinfo {author} {\bibnamefont {Ren}, \bibfnamefont
  {Z.}}, \bibinfo {author} {\bibfnamefont {A.~A.}\ \bibnamefont {Taskin}},
  \bibinfo {author} {\bibfnamefont {S.}~\bibnamefont {Sasaki}}, \bibinfo
  {author} {\bibfnamefont {K.}~\bibnamefont {Segawa}}, \ and\ \bibinfo {author}
  {\bibfnamefont {Y.}~\bibnamefont {Ando}}} (\bibinfo {year} {2011}),\ \href
  {\doibase 10.1103/PhysRevB.84.165311} {\bibfield  {journal} {\bibinfo
  {journal} {Phys. Rev. B}\ }\textbf {\bibinfo {volume} {84}}~(\bibinfo
  {number} {16}),\ \bibinfo {pages} {165311}}\BibitemShut {NoStop}%
\bibitem [{\citenamefont {Roy}(2009)}]{Roy2009}%
  \BibitemOpen
  \bibfield  {author} {\bibinfo {author} {\bibnamefont {Roy}, \bibfnamefont
  {R.}}} (\bibinfo {year} {2009}),\ \href@noop {} {\bibfield  {journal}
  {\bibinfo  {journal} {Phys. Rev. B}\ }\textbf {\bibinfo {volume}
  {79}}~(\bibinfo {number} {19}),\ \bibinfo {pages} {195322}}\BibitemShut
  {NoStop}%
\bibitem [{\citenamefont {Sato}\ \emph {et~al.}(2011)\citenamefont {Sato},
  \citenamefont {Segawa}, \citenamefont {Kosaka}, \citenamefont {Souma},
  \citenamefont {Nakayama}, \citenamefont {Eto}, \citenamefont {Minami},
  \citenamefont {Ando},\ and\ \citenamefont {Takahashi}}]{Sato2011}%
  \BibitemOpen
  \bibfield  {author} {\bibinfo {author} {\bibnamefont {Sato}, \bibfnamefont
  {T.}}, \bibinfo {author} {\bibfnamefont {K.}~\bibnamefont {Segawa}}, \bibinfo
  {author} {\bibfnamefont {K.}~\bibnamefont {Kosaka}}, \bibinfo {author}
  {\bibfnamefont {S.}~\bibnamefont {Souma}}, \bibinfo {author} {\bibfnamefont
  {K.}~\bibnamefont {Nakayama}}, \bibinfo {author} {\bibfnamefont
  {K.}~\bibnamefont {Eto}}, \bibinfo {author} {\bibfnamefont {T.}~\bibnamefont
  {Minami}}, \bibinfo {author} {\bibfnamefont {Y.}~\bibnamefont {Ando}}, \ and\
  \bibinfo {author} {\bibfnamefont {T.}~\bibnamefont {Takahashi}}} (\bibinfo
  {year} {2011}),\ \href {\doibase 10.1038/nphys2058} {\bibfield  {journal}
  {\bibinfo  {journal} {Nat. Phys.}\ }\textbf {\bibinfo {volume} {7}}~(\bibinfo
  {number} {11}),\ \bibinfo {pages} {840}}\BibitemShut {NoStop}%
\bibitem [{\citenamefont {Schafgans}\ \emph {et~al.}(2012)\citenamefont
  {Schafgans}, \citenamefont {Post}, \citenamefont {Taskin}, \citenamefont
  {Ando}, \citenamefont {Qi}, \citenamefont {Chapler},\ and\ \citenamefont
  {Basov}}]{Schafgans2012}%
  \BibitemOpen
  \bibfield  {author} {\bibinfo {author} {\bibnamefont {Schafgans},
  \bibfnamefont {A.~A.}}, \bibinfo {author} {\bibfnamefont {K.~W.}\
  \bibnamefont {Post}}, \bibinfo {author} {\bibfnamefont {A.~A.}\ \bibnamefont
  {Taskin}}, \bibinfo {author} {\bibfnamefont {Y.}~\bibnamefont {Ando}},
  \bibinfo {author} {\bibfnamefont {X.-L.}\ \bibnamefont {Qi}}, \bibinfo
  {author} {\bibfnamefont {B.~C.}\ \bibnamefont {Chapler}}, \ and\ \bibinfo
  {author} {\bibfnamefont {D.~N.}\ \bibnamefont {Basov}}} (\bibinfo {year}
  {2012}),\ \href {\doibase 10.1103/PhysRevB.85.195440} {\bibfield  {journal}
  {\bibinfo  {journal} {Phys. Rev. B}\ }\textbf {\bibinfo {volume}
  {85}}~(\bibinfo {number} {19}),\ \bibinfo {pages} {195440}}\BibitemShut
  {NoStop}%
\bibitem [{\citenamefont {Schindler}\ \emph
  {et~al.}(2018{\natexlab{a}})\citenamefont {Schindler}, \citenamefont {Cook},
  \citenamefont {Vergniory}, \citenamefont {Wang}, \citenamefont {Parkin},
  \citenamefont {Bernevig},\ and\ \citenamefont {Neupert}}]{Schindler2018SA}%
  \BibitemOpen
  \bibfield  {author} {\bibinfo {author} {\bibnamefont {Schindler},
  \bibfnamefont {F.}}, \bibinfo {author} {\bibfnamefont {A.~M.}\ \bibnamefont
  {Cook}}, \bibinfo {author} {\bibfnamefont {M.~G.}\ \bibnamefont {Vergniory}},
  \bibinfo {author} {\bibfnamefont {Z.}~\bibnamefont {Wang}}, \bibinfo {author}
  {\bibfnamefont {S.~S.~P.}\ \bibnamefont {Parkin}}, \bibinfo {author}
  {\bibfnamefont {B.~A.}\ \bibnamefont {Bernevig}}, \ and\ \bibinfo {author}
  {\bibfnamefont {T.}~\bibnamefont {Neupert}}} (\bibinfo {year}
  {2018}{\natexlab{a}}),\ \href {\doibase 10.1126/sciadv.aat0346} {\bibfield
  {journal} {\bibinfo  {journal} {Science Advances}\ }\textbf {\bibinfo
  {volume} {4}}~(\bibinfo {number} {6}),\ \bibinfo {pages}
  {eaat0346}}\BibitemShut {NoStop}%
\bibitem [{\citenamefont {Schindler}\ \emph
  {et~al.}(2018{\natexlab{b}})\citenamefont {Schindler}, \citenamefont {Wang},
  \citenamefont {Vergniory}, \citenamefont {Cook}, \citenamefont {Murani},
  \citenamefont {Sengupta}, \citenamefont {Kasumov}, \citenamefont {Deblock},
  \citenamefont {Jeon}, \citenamefont {Drozdov}, \citenamefont {Bouchiat},
  \citenamefont {Guéron}, \citenamefont {Yazdani}, \citenamefont {Bernevig},\
  and\ \citenamefont {Neupert}}]{Schindler2018NP}%
  \BibitemOpen
  \bibfield  {author} {\bibinfo {author} {\bibnamefont {Schindler},
  \bibfnamefont {F.}}, \bibinfo {author} {\bibfnamefont {Z.}~\bibnamefont
  {Wang}}, \bibinfo {author} {\bibfnamefont {M.~G.}\ \bibnamefont {Vergniory}},
  \bibinfo {author} {\bibfnamefont {A.~M.}\ \bibnamefont {Cook}}, \bibinfo
  {author} {\bibfnamefont {A.}~\bibnamefont {Murani}}, \bibinfo {author}
  {\bibfnamefont {S.}~\bibnamefont {Sengupta}}, \bibinfo {author}
  {\bibfnamefont {A.~Y.}\ \bibnamefont {Kasumov}}, \bibinfo {author}
  {\bibfnamefont {R.}~\bibnamefont {Deblock}}, \bibinfo {author} {\bibfnamefont
  {S.}~\bibnamefont {Jeon}}, \bibinfo {author} {\bibfnamefont {I.}~\bibnamefont
  {Drozdov}}, \bibinfo {author} {\bibfnamefont {H.}~\bibnamefont {Bouchiat}},
  \bibinfo {author} {\bibfnamefont {S.}~\bibnamefont {Guéron}}, \bibinfo
  {author} {\bibfnamefont {A.}~\bibnamefont {Yazdani}}, \bibinfo {author}
  {\bibfnamefont {B.~A.}\ \bibnamefont {Bernevig}}, \ and\ \bibinfo {author}
  {\bibfnamefont {T.}~\bibnamefont {Neupert}}} (\bibinfo {year}
  {2018}{\natexlab{b}}),\ \href {\doibase 10.1038/s41567-018-0224-7} {\bibfield
   {journal} {\bibinfo  {journal} {Nature Physics}\ }\textbf {\bibinfo {volume}
  {14}}~(\bibinfo {number} {9}),\ \bibinfo {pages} {918}}\BibitemShut {NoStop}%
\bibitem [{\citenamefont {Sheng}\ \emph {et~al.}(2006)\citenamefont {Sheng},
  \citenamefont {Weng}, \citenamefont {Sheng},\ and\ \citenamefont
  {Haldane}}]{Sheng2006}%
  \BibitemOpen
  \bibfield  {author} {\bibinfo {author} {\bibnamefont {Sheng}, \bibfnamefont
  {D.~N.}}, \bibinfo {author} {\bibfnamefont {Z.~Y.}\ \bibnamefont {Weng}},
  \bibinfo {author} {\bibfnamefont {L.}~\bibnamefont {Sheng}}, \ and\ \bibinfo
  {author} {\bibfnamefont {F.~D.~M.}\ \bibnamefont {Haldane}}} (\bibinfo {year}
  {2006}),\ \href@noop {} {\bibfield  {journal} {\bibinfo  {journal} {Phys.
  Rev. Lett.}\ }\textbf {\bibinfo {volume} {97}},\ \bibinfo {pages}
  {036808}}\BibitemShut {NoStop}%
\bibitem [{\citenamefont {Shojaei}\ \emph {et~al.}(2018)\citenamefont
  {Shojaei}, \citenamefont {McFadden}, \citenamefont {Pendharkar},
  \citenamefont {Lee}, \citenamefont {Flatté},\ and\ \citenamefont
  {Palmstrøm}}]{Shojaei2018}%
  \BibitemOpen
  \bibfield  {author} {\bibinfo {author} {\bibnamefont {Shojaei}, \bibfnamefont
  {B.}}, \bibinfo {author} {\bibfnamefont {A.~P.}\ \bibnamefont {McFadden}},
  \bibinfo {author} {\bibfnamefont {M.}~\bibnamefont {Pendharkar}}, \bibinfo
  {author} {\bibfnamefont {J.~S.}\ \bibnamefont {Lee}}, \bibinfo {author}
  {\bibfnamefont {M.~E.}\ \bibnamefont {Flatté}}, \ and\ \bibinfo {author}
  {\bibfnamefont {C.~J.}\ \bibnamefont {Palmstrøm}}} (\bibinfo {year}
  {2018}),\ \href {\doibase 10.1103/PhysRevMaterials.2.064603} {\bibfield
  {journal} {\bibinfo  {journal} {Phys. Rev. Mater.}\ }\textbf {\bibinfo
  {volume} {2}}~(\bibinfo {number} {6}),\ \bibinfo {pages}
  {064603}}\BibitemShut {NoStop}%
\bibitem [{\citenamefont {Tanaka}\ \emph {et~al.}(2012)\citenamefont {Tanaka},
  \citenamefont {Ren}, \citenamefont {Sato}, \citenamefont {Nakayama},
  \citenamefont {Souma}, \citenamefont {Takahashi}, \citenamefont {Segawa},\
  and\ \citenamefont {Ando}}]{Tanaka2012}%
  \BibitemOpen
  \bibfield  {author} {\bibinfo {author} {\bibnamefont {Tanaka}, \bibfnamefont
  {Y.}}, \bibinfo {author} {\bibfnamefont {Z.}~\bibnamefont {Ren}}, \bibinfo
  {author} {\bibfnamefont {T.}~\bibnamefont {Sato}}, \bibinfo {author}
  {\bibfnamefont {K.}~\bibnamefont {Nakayama}}, \bibinfo {author}
  {\bibfnamefont {S.}~\bibnamefont {Souma}}, \bibinfo {author} {\bibfnamefont
  {T.}~\bibnamefont {Takahashi}}, \bibinfo {author} {\bibfnamefont
  {K.}~\bibnamefont {Segawa}}, \ and\ \bibinfo {author} {\bibfnamefont
  {Y.}~\bibnamefont {Ando}}} (\bibinfo {year} {2012}),\ \href@noop {}
  {\bibfield  {journal} {\bibinfo  {journal} {Nat. Phys.}\ }\textbf {\bibinfo
  {volume} {8}}~(\bibinfo {number} {11}),\ \bibinfo {pages} {800}}\BibitemShut
  {NoStop}%
\bibitem [{\citenamefont {Tanaka}\ \emph
  {et~al.}(2013{\natexlab{a}})\citenamefont {Tanaka}, \citenamefont {Sato},
  \citenamefont {Nakayama}, \citenamefont {Souma}, \citenamefont {Takahashi},
  \citenamefont {Ren}, \citenamefont {Novak}, \citenamefont {Segawa},\ and\
  \citenamefont {Ando}}]{Tanaka2013b}%
  \BibitemOpen
  \bibfield  {author} {\bibinfo {author} {\bibnamefont {Tanaka}, \bibfnamefont
  {Y.}}, \bibinfo {author} {\bibfnamefont {T.}~\bibnamefont {Sato}}, \bibinfo
  {author} {\bibfnamefont {K.}~\bibnamefont {Nakayama}}, \bibinfo {author}
  {\bibfnamefont {S.}~\bibnamefont {Souma}}, \bibinfo {author} {\bibfnamefont
  {T.}~\bibnamefont {Takahashi}}, \bibinfo {author} {\bibfnamefont
  {Z.}~\bibnamefont {Ren}}, \bibinfo {author} {\bibfnamefont {M.}~\bibnamefont
  {Novak}}, \bibinfo {author} {\bibfnamefont {K.}~\bibnamefont {Segawa}}, \
  and\ \bibinfo {author} {\bibfnamefont {Y.}~\bibnamefont {Ando}}} (\bibinfo
  {year} {2013}{\natexlab{a}}),\ \href {\doibase 10.1103/PhysRevB.87.155105}
  {\bibfield  {journal} {\bibinfo  {journal} {Phys. Rev. B}\ }\textbf {\bibinfo
  {volume} {87}}~(\bibinfo {number} {15}),\ \bibinfo {pages}
  {155105}}\BibitemShut {NoStop}%
\bibitem [{\citenamefont {Tanaka}\ \emph
  {et~al.}(2013{\natexlab{b}})\citenamefont {Tanaka}, \citenamefont {Shoman},
  \citenamefont {Nakayama}, \citenamefont {Souma}, \citenamefont {Sato},
  \citenamefont {Takahashi}, \citenamefont {Novak}, \citenamefont {Segawa},\
  and\ \citenamefont {Ando}}]{Tanaka2013a}%
  \BibitemOpen
  \bibfield  {author} {\bibinfo {author} {\bibnamefont {Tanaka}, \bibfnamefont
  {Y.}}, \bibinfo {author} {\bibfnamefont {T.}~\bibnamefont {Shoman}}, \bibinfo
  {author} {\bibfnamefont {K.}~\bibnamefont {Nakayama}}, \bibinfo {author}
  {\bibfnamefont {S.}~\bibnamefont {Souma}}, \bibinfo {author} {\bibfnamefont
  {T.}~\bibnamefont {Sato}}, \bibinfo {author} {\bibfnamefont {T.}~\bibnamefont
  {Takahashi}}, \bibinfo {author} {\bibfnamefont {M.}~\bibnamefont {Novak}},
  \bibinfo {author} {\bibfnamefont {K.}~\bibnamefont {Segawa}}, \ and\ \bibinfo
  {author} {\bibfnamefont {Y.}~\bibnamefont {Ando}}} (\bibinfo {year}
  {2013}{\natexlab{b}}),\ \href@noop {} {\bibfield  {journal} {\bibinfo
  {journal} {Phys. Rev. B}\ }\textbf {\bibinfo {volume} {88}}~(\bibinfo
  {number} {23}),\ \bibinfo {pages} {235126}}\BibitemShut {NoStop}%
\bibitem [{\citenamefont {Taskin}\ and\ \citenamefont
  {Ando}(2009)}]{Taskin2009}%
  \BibitemOpen
  \bibfield  {author} {\bibinfo {author} {\bibnamefont {Taskin}, \bibfnamefont
  {A.~A.}}, \ and\ \bibinfo {author} {\bibfnamefont {Y.}~\bibnamefont {Ando}}}
  (\bibinfo {year} {2009}),\ \href {\doibase 10.1103/PhysRevB.80.085303}
  {\bibfield  {journal} {\bibinfo  {journal} {Phys. Rev. B}\ }\textbf {\bibinfo
  {volume} {80}}~(\bibinfo {number} {8}),\ \bibinfo {pages}
  {085303}}\BibitemShut {NoStop}%
\bibitem [{\citenamefont {Taskin}\ and\ \citenamefont
  {Ando}(2011)}]{Taskin2011b}%
  \BibitemOpen
  \bibfield  {author} {\bibinfo {author} {\bibnamefont {Taskin}, \bibfnamefont
  {A.~A.}}, \ and\ \bibinfo {author} {\bibfnamefont {Y.}~\bibnamefont {Ando}}}
  (\bibinfo {year} {2011}),\ \href {\doibase 10.1103/PhysRevB.84.035301}
  {\bibfield  {journal} {\bibinfo  {journal} {Phys. Rev. B}\ }\textbf {\bibinfo
  {volume} {84}}~(\bibinfo {number} {3}),\ \bibinfo {pages}
  {035301}}\BibitemShut {NoStop}%
\bibitem [{\citenamefont {Taskin}\ \emph {et~al.}(2011)\citenamefont {Taskin},
  \citenamefont {Ren}, \citenamefont {Sasaki}, \citenamefont {Segawa},\ and\
  \citenamefont {Ando}}]{Taskin2011}%
  \BibitemOpen
  \bibfield  {author} {\bibinfo {author} {\bibnamefont {Taskin}, \bibfnamefont
  {A.~A.}}, \bibinfo {author} {\bibfnamefont {Z.}~\bibnamefont {Ren}}, \bibinfo
  {author} {\bibfnamefont {S.}~\bibnamefont {Sasaki}}, \bibinfo {author}
  {\bibfnamefont {K.}~\bibnamefont {Segawa}}, \ and\ \bibinfo {author}
  {\bibfnamefont {Y.}~\bibnamefont {Ando}}} (\bibinfo {year} {2011}),\ \href
  {\doibase 10.1103/PhysRevLett.107.016801} {\bibfield  {journal} {\bibinfo
  {journal} {Phys. Rev. Lett.}\ }\textbf {\bibinfo {volume} {107}}~(\bibinfo
  {number} {1}),\ \bibinfo {pages} {016801}}\BibitemShut {NoStop}%
\bibitem [{\citenamefont {Teo}\ \emph {et~al.}(2008)\citenamefont {Teo},
  \citenamefont {Fu},\ and\ \citenamefont {Kane}}]{Teo2008}%
  \BibitemOpen
  \bibfield  {author} {\bibinfo {author} {\bibnamefont {Teo}, \bibfnamefont
  {J.~C.~Y.}}, \bibinfo {author} {\bibfnamefont {L.}~\bibnamefont {Fu}}, \ and\
  \bibinfo {author} {\bibfnamefont {C.~L.}\ \bibnamefont {Kane}}} (\bibinfo
  {year} {2008}),\ \href {\doibase 10.1103/PhysRevB.78.045426} {\bibfield
  {journal} {\bibinfo  {journal} {Phys. Rev. B}\ }\textbf {\bibinfo {volume}
  {78}}~(\bibinfo {number} {4}),\ \bibinfo {pages} {045426}}\BibitemShut
  {NoStop}%
\bibitem [{\citenamefont {Thouless}\ \emph {et~al.}(1982)\citenamefont
  {Thouless}, \citenamefont {Kohmoto}, \citenamefont {Nightingale},\ and\
  \citenamefont {Dennijs}}]{Thouless1982}%
  \BibitemOpen
  \bibfield  {author} {\bibinfo {author} {\bibnamefont {Thouless},
  \bibfnamefont {D.~J.}}, \bibinfo {author} {\bibfnamefont {M.}~\bibnamefont
  {Kohmoto}}, \bibinfo {author} {\bibfnamefont {M.~P.}\ \bibnamefont
  {Nightingale}}, \ and\ \bibinfo {author} {\bibfnamefont {M.}~\bibnamefont
  {Dennijs}}} (\bibinfo {year} {1982}),\ \href@noop {} {\bibfield  {journal}
  {\bibinfo  {journal} {Phys. Rev. Lett.}\ }\textbf {\bibinfo {volume}
  {49}}~(\bibinfo {number} {6}),\ \bibinfo {pages} {405}}\BibitemShut {NoStop}%
\bibitem [{\citenamefont {Wu}\ \emph {et~al.}(2018)\citenamefont {Wu},
  \citenamefont {Fatemi}, \citenamefont {Gibson~Quinn}, \citenamefont
  {Watanabe}, \citenamefont {Taniguchi}, \citenamefont {Cava~Robert},\ and\
  \citenamefont {Jarillo-Herrero}}]{Wu2018}%
  \BibitemOpen
  \bibfield  {author} {\bibinfo {author} {\bibnamefont {Wu}, \bibfnamefont
  {S.}}, \bibinfo {author} {\bibfnamefont {V.}~\bibnamefont {Fatemi}}, \bibinfo
  {author} {\bibfnamefont {D.}~\bibnamefont {Gibson~Quinn}}, \bibinfo {author}
  {\bibfnamefont {K.}~\bibnamefont {Watanabe}}, \bibinfo {author}
  {\bibfnamefont {T.}~\bibnamefont {Taniguchi}}, \bibinfo {author}
  {\bibfnamefont {J.}~\bibnamefont {Cava~Robert}}, \ and\ \bibinfo {author}
  {\bibfnamefont {P.}~\bibnamefont {Jarillo-Herrero}}} (\bibinfo {year}
  {2018}),\ \href {\doibase 10.1126/science.aan6003} {\bibfield  {journal}
  {\bibinfo  {journal} {Science}\ }\textbf {\bibinfo {volume} {359}}~(\bibinfo
  {number} {6371}),\ \bibinfo {pages} {76}}\BibitemShut {NoStop}%
\bibitem [{\citenamefont {Xia}\ \emph {et~al.}(2009)\citenamefont {Xia},
  \citenamefont {Qian}, \citenamefont {Hsieh}, \citenamefont {Wray},
  \citenamefont {Pal}, \citenamefont {Lin}, \citenamefont {Bansil},
  \citenamefont {Grauer}, \citenamefont {Hor}, \citenamefont {Cava},\ and\
  \citenamefont {Hasan}}]{Xia2009}%
  \BibitemOpen
  \bibfield  {author} {\bibinfo {author} {\bibnamefont {Xia}, \bibfnamefont
  {Y.}}, \bibinfo {author} {\bibfnamefont {D.}~\bibnamefont {Qian}}, \bibinfo
  {author} {\bibfnamefont {D.}~\bibnamefont {Hsieh}}, \bibinfo {author}
  {\bibfnamefont {L.}~\bibnamefont {Wray}}, \bibinfo {author} {\bibfnamefont
  {A.}~\bibnamefont {Pal}}, \bibinfo {author} {\bibfnamefont {H.}~\bibnamefont
  {Lin}}, \bibinfo {author} {\bibfnamefont {A.}~\bibnamefont {Bansil}},
  \bibinfo {author} {\bibfnamefont {D.}~\bibnamefont {Grauer}}, \bibinfo
  {author} {\bibfnamefont {Y.~S.}\ \bibnamefont {Hor}}, \bibinfo {author}
  {\bibfnamefont {R.~J.}\ \bibnamefont {Cava}}, \ and\ \bibinfo {author}
  {\bibfnamefont {M.~Z.}\ \bibnamefont {Hasan}}} (\bibinfo {year} {2009}),\
  \href {\doibase 10.1038/nphys1274} {\bibfield  {journal} {\bibinfo  {journal}
  {Nat. Phys.}\ }\textbf {\bibinfo {volume} {5}}~(\bibinfo {number} {6}),\
  \bibinfo {pages} {398}}\BibitemShut {NoStop}%
\bibitem [{\citenamefont {Xu}\ \emph {et~al.}(2012)\citenamefont {Xu},
  \citenamefont {Liu}, \citenamefont {Alidoust}, \citenamefont {Neupane},
  \citenamefont {Qian}, \citenamefont {Belopolski}, \citenamefont {Denlinger},
  \citenamefont {Wang}, \citenamefont {Lin}, \citenamefont {Wray},
  \citenamefont {Landolt}, \citenamefont {Slomski}, \citenamefont {Dil},
  \citenamefont {Marcinkova}, \citenamefont {Morosan}, \citenamefont {Gibson},
  \citenamefont {Sankar}, \citenamefont {Chou}, \citenamefont {Cava},
  \citenamefont {Bansil},\ and\ \citenamefont {Hasan}}]{Xu2012}%
  \BibitemOpen
  \bibfield  {author} {\bibinfo {author} {\bibnamefont {Xu}, \bibfnamefont
  {S.-Y.}}, \bibinfo {author} {\bibfnamefont {C.}~\bibnamefont {Liu}}, \bibinfo
  {author} {\bibfnamefont {N.}~\bibnamefont {Alidoust}}, \bibinfo {author}
  {\bibfnamefont {M.}~\bibnamefont {Neupane}}, \bibinfo {author} {\bibfnamefont
  {D.}~\bibnamefont {Qian}}, \bibinfo {author} {\bibfnamefont {I.}~\bibnamefont
  {Belopolski}}, \bibinfo {author} {\bibfnamefont {J.~D.}\ \bibnamefont
  {Denlinger}}, \bibinfo {author} {\bibfnamefont {Y.~J.}\ \bibnamefont {Wang}},
  \bibinfo {author} {\bibfnamefont {H.}~\bibnamefont {Lin}}, \bibinfo {author}
  {\bibfnamefont {L.~A.}\ \bibnamefont {Wray}}, \bibinfo {author}
  {\bibfnamefont {G.}~\bibnamefont {Landolt}}, \bibinfo {author} {\bibfnamefont
  {B.}~\bibnamefont {Slomski}}, \bibinfo {author} {\bibfnamefont {J.~H.}\
  \bibnamefont {Dil}}, \bibinfo {author} {\bibfnamefont {A.}~\bibnamefont
  {Marcinkova}}, \bibinfo {author} {\bibfnamefont {E.}~\bibnamefont {Morosan}},
  \bibinfo {author} {\bibfnamefont {Q.}~\bibnamefont {Gibson}}, \bibinfo
  {author} {\bibfnamefont {R.}~\bibnamefont {Sankar}}, \bibinfo {author}
  {\bibfnamefont {F.~C.}\ \bibnamefont {Chou}}, \bibinfo {author}
  {\bibfnamefont {R.~J.}\ \bibnamefont {Cava}}, \bibinfo {author}
  {\bibfnamefont {A.}~\bibnamefont {Bansil}}, \ and\ \bibinfo {author}
  {\bibfnamefont {M.~Z.}\ \bibnamefont {Hasan}}} (\bibinfo {year} {2012}),\
  \href {\doibase 10.1038/ncomms2191} {\bibfield  {journal} {\bibinfo
  {journal} {Nat. Commun.}\ }\textbf {\bibinfo {volume} {3}},\ \bibinfo {pages}
  {1192}}\BibitemShut {NoStop}%
\bibitem [{\citenamefont {Xu}\ \emph {et~al.}(2011)\citenamefont {Xu},
  \citenamefont {Xia}, \citenamefont {Wray}, \citenamefont {Jia}, \citenamefont
  {Meier}, \citenamefont {Dil}, \citenamefont {Osterwalder}, \citenamefont
  {Slomski}, \citenamefont {Bansil}, \citenamefont {Lin}, \citenamefont
  {Cava},\ and\ \citenamefont {Hasan}}]{Xu2011}%
  \BibitemOpen
  \bibfield  {author} {\bibinfo {author} {\bibnamefont {Xu}, \bibfnamefont
  {S.-Y.}}, \bibinfo {author} {\bibfnamefont {Y.}~\bibnamefont {Xia}}, \bibinfo
  {author} {\bibfnamefont {L.~A.}\ \bibnamefont {Wray}}, \bibinfo {author}
  {\bibfnamefont {S.}~\bibnamefont {Jia}}, \bibinfo {author} {\bibfnamefont
  {F.}~\bibnamefont {Meier}}, \bibinfo {author} {\bibfnamefont {J.~H.}\
  \bibnamefont {Dil}}, \bibinfo {author} {\bibfnamefont {J.}~\bibnamefont
  {Osterwalder}}, \bibinfo {author} {\bibfnamefont {B.}~\bibnamefont
  {Slomski}}, \bibinfo {author} {\bibfnamefont {A.}~\bibnamefont {Bansil}},
  \bibinfo {author} {\bibfnamefont {H.}~\bibnamefont {Lin}}, \bibinfo {author}
  {\bibfnamefont {R.~J.}\ \bibnamefont {Cava}}, \ and\ \bibinfo {author}
  {\bibfnamefont {M.~Z.}\ \bibnamefont {Hasan}}} (\bibinfo {year} {2011}),\
  \href {\doibase 10.1126/science.1201607} {\bibfield  {journal} {\bibinfo
  {journal} {Science}\ }\textbf {\bibinfo {volume} {332}}~(\bibinfo {number}
  {6029}),\ \bibinfo {pages} {560}}\BibitemShut {NoStop}%
\bibitem [{\citenamefont {Yonezawa}(2019)}]{Yonezawa2019}%
  \BibitemOpen
  \bibfield  {author} {\bibinfo {author} {\bibnamefont {Yonezawa},
  \bibfnamefont {S.}}} (\bibinfo {year} {2019}),\ \href@noop {} {\bibfield
  {journal} {\bibinfo  {journal} {Cond. Matter}\ }\textbf {\bibinfo {volume}
  {4}}~(\bibinfo {number} {1}),\ \bibinfo {pages} {2}}\BibitemShut {NoStop}%
\bibitem [{\citenamefont {Zhang}\ \emph {et~al.}(2009)\citenamefont {Zhang},
  \citenamefont {Liu}, \citenamefont {Qi}, \citenamefont {Dai}, \citenamefont
  {Fang},\ and\ \citenamefont {Zhang}}]{Zhang2009}%
  \BibitemOpen
  \bibfield  {author} {\bibinfo {author} {\bibnamefont {Zhang}, \bibfnamefont
  {H.}}, \bibinfo {author} {\bibfnamefont {C.-X.}\ \bibnamefont {Liu}},
  \bibinfo {author} {\bibfnamefont {X.~L.}\ \bibnamefont {Qi}}, \bibinfo
  {author} {\bibfnamefont {X.}~\bibnamefont {Dai}}, \bibinfo {author}
  {\bibfnamefont {Z.}~\bibnamefont {Fang}}, \ and\ \bibinfo {author}
  {\bibfnamefont {S.-C.}\ \bibnamefont {Zhang}}} (\bibinfo {year} {2009}),\
  \href@noop {} {\bibfield  {journal} {\bibinfo  {journal} {Nat. Phys.}\
  }\textbf {\bibinfo {volume} {5}},\ \bibinfo {pages} {438}}\BibitemShut
  {NoStop}%
\bibitem [{\citenamefont {Zhang}\ \emph {et~al.}(2011)\citenamefont {Zhang},
  \citenamefont {Chang}, \citenamefont {Zhang}, \citenamefont {Wen},
  \citenamefont {Feng}, \citenamefont {Li}, \citenamefont {Liu}, \citenamefont
  {He}, \citenamefont {Wang}, \citenamefont {Chen}, \citenamefont {Xue},
  \citenamefont {Ma},\ and\ \citenamefont {Wang}}]{Zhang2011a}%
  \BibitemOpen
  \bibfield  {author} {\bibinfo {author} {\bibnamefont {Zhang}, \bibfnamefont
  {J.}}, \bibinfo {author} {\bibfnamefont {C.-Z.}\ \bibnamefont {Chang}},
  \bibinfo {author} {\bibfnamefont {Z.}~\bibnamefont {Zhang}}, \bibinfo
  {author} {\bibfnamefont {J.}~\bibnamefont {Wen}}, \bibinfo {author}
  {\bibfnamefont {X.}~\bibnamefont {Feng}}, \bibinfo {author} {\bibfnamefont
  {K.}~\bibnamefont {Li}}, \bibinfo {author} {\bibfnamefont {M.}~\bibnamefont
  {Liu}}, \bibinfo {author} {\bibfnamefont {K.}~\bibnamefont {He}}, \bibinfo
  {author} {\bibfnamefont {L.}~\bibnamefont {Wang}}, \bibinfo {author}
  {\bibfnamefont {X.}~\bibnamefont {Chen}}, \bibinfo {author} {\bibfnamefont
  {Q.-K.}\ \bibnamefont {Xue}}, \bibinfo {author} {\bibfnamefont
  {X.}~\bibnamefont {Ma}}, \ and\ \bibinfo {author} {\bibfnamefont
  {Y.}~\bibnamefont {Wang}}} (\bibinfo {year} {2011}),\ \href {\doibase
  10.1038/ncomms1588} {\bibfield  {journal} {\bibinfo  {journal} {Nat.
  Commun.}\ }\textbf {\bibinfo {volume} {2}},\ \bibinfo {pages}
  {574}}\BibitemShut {NoStop}%
\bibitem [{\citenamefont {Zhang}\ and\ \citenamefont {Hu}(2001)}]{Zhang2001}%
  \BibitemOpen
  \bibfield  {author} {\bibinfo {author} {\bibnamefont {Zhang}, \bibfnamefont
  {S.~C.}}, \ and\ \bibinfo {author} {\bibfnamefont {J.~P.}\ \bibnamefont
  {Hu}}} (\bibinfo {year} {2001}),\ \href {\doibase
  10.1126/science.294.5543.823} {\bibfield  {journal} {\bibinfo  {journal}
  {Science}\ }\textbf {\bibinfo {volume} {294}}~(\bibinfo {number} {5543}),\
  \bibinfo {pages} {823}}\BibitemShut {NoStop}%
\end{thebibliography}

%merlin.mbs apsrmp4-1.bst 2010-07-25 4.21a (PWD, AO, DPC) hacked
%Control: author (11) reversed first initials

\end{document}